\documentclass[twocolumn,superscriptaddress,10pt]{revtex4-1}
\usepackage{graphicx}
\usepackage{epsfig}
\usepackage{amsmath}
\usepackage{amssymb}
\usepackage{amsfonts}
\usepackage{mathrsfs}
\usepackage{theorem}
\usepackage{bm}
\usepackage{url}
\usepackage[T1]{fontenc}
\usepackage{csquotes}
\usepackage{natbib}
\usepackage{float}
\usepackage{appendix}
\usepackage{color}
\MakeOuterQuote{"}
\usepackage{dcolumn}
\usepackage{color}

\begin{document}
\title{Experimental high-dimensional quantum teleportation}
\author{Xiao-Min Hu}
\email{These two authors contributed equally to this work.}
\affiliation{CAS Key Laboratory of Quantum Information, University of Science and Technology of China, Hefei, 230026, People's Republic of China}
\affiliation{CAS Center For Excellence in Quantum Information and Quantum Physics, University of Science and Technology of China, Hefei, 230026, People's Republic of China}

\author{Chao Zhang}
\email{These two authors contributed equally to this work.}
\affiliation{CAS Key Laboratory of Quantum Information, University of Science and Technology of China, Hefei, 230026, People's Republic of China}
\affiliation{CAS Center For Excellence in Quantum Information and Quantum Physics, University of Science and Technology of China, Hefei, 230026, People's Republic of China}

\author{Bi-Heng Liu}
\email{bhliu@ustc.edu.cn}
\affiliation{CAS Key Laboratory of Quantum Information, University of Science and Technology of China, Hefei, 230026, People's Republic of China}
\affiliation{CAS Center For Excellence in Quantum Information and Quantum Physics, University of Science and Technology of China, Hefei, 230026, People's Republic of China}

\author{Yu Cai}
\affiliation {Department of Applied Physics, University of Geneva, CH-1211 Geneva, Switzerland}

\author{Xiang-Jun Ye}
\affiliation{CAS Key Laboratory of Quantum Information, University of Science and Technology of China, Hefei, 230026, People's Republic of China}
\affiliation{CAS Center For Excellence in Quantum Information and Quantum Physics, University of Science and Technology of China, Hefei, 230026, People's Republic of China}

\author{Yu Guo}
\affiliation{CAS Key Laboratory of Quantum Information, University of Science and Technology of China, Hefei, 230026, People's Republic of China}
\affiliation{CAS Center For Excellence in Quantum Information and Quantum Physics, University of Science and Technology of China, Hefei, 230026, People's Republic of China}

\author{Wen-Bo Xing}
\affiliation{CAS Key Laboratory of Quantum Information, University of Science and Technology of China, Hefei, 230026, People's Republic of China}
\affiliation{CAS Center For Excellence in Quantum Information and Quantum Physics, University of Science and Technology of China, Hefei, 230026, People's Republic of China}

\author{Cen-Xiao Huang}
\affiliation{CAS Key Laboratory of Quantum Information, University of Science and Technology of China, Hefei, 230026, People's Republic of China}
\affiliation{CAS Center For Excellence in Quantum Information and Quantum Physics, University of Science and Technology of China, Hefei, 230026, People's Republic of China}

\author{Yun-Feng Huang}
\affiliation{CAS Key Laboratory of Quantum Information, University of Science and Technology of China, Hefei, 230026, People's Republic of China}
\affiliation{CAS Center For Excellence in Quantum Information and Quantum Physics, University of Science and Technology of China, Hefei, 230026, People's Republic of China}

\author{Chuan-Feng Li}
\email{cfli@ustc.edu.cn}
\affiliation{CAS Key Laboratory of Quantum Information, University of Science and Technology of China, Hefei, 230026, People's Republic of China}
\affiliation{CAS Center For Excellence in Quantum Information and Quantum Physics, University of Science and Technology of China, Hefei, 230026, People's Republic of China}

\author{Guang-Can Guo}
\affiliation{CAS Key Laboratory of Quantum Information, University of Science and Technology of China, Hefei, 230026, People's Republic of China}
\affiliation{CAS Center For Excellence in Quantum Information and Quantum Physics, University of Science and Technology of China, Hefei, 230026, People's Republic of China}

\begin{abstract}
Quantum teleportation provides a way to transmit unknown quantum states from one location to another. In the quantum world, multilevel systems which enable high-dimensional systems are more prevalent. Therefore, to completely rebuild the quantum states of a single particle remotely, one needs to teleport multilevel (high-dimensional) states. Here, we demonstrate the teleportation of high-dimensional states in a three-dimensional six-photon system. We exploit the spatial mode of a single photon as the high-dimensional system, use two auxiliary entangled photons to realize a deterministic three-dimensional Bell state measurement. The fidelity of teleportaion process matrix is $F=0.596\pm0.037$. Through this process matrix, we can prove that our teleportation is both non-classical and genuine three-dimensional. Our work paves the way to rebuild complex quantum systems remotely and to construct complex quantum networks.
\end{abstract}
\maketitle

Quantum teleportation\cite{Bennett93,Pirandola15} enables the rebuilding of arbitrary unknown quantum states without the transmission of a real particle. Previous efforts have shown the capability to rebuild qubit states and continuous variable states. Discrete variable states\cite{Bouwmeester97,Nielsen98,Fattal04,Barrett04,Riebe04,Sherson06,Olmschenk09} and continuous variable states \cite{Furusawa98,Takei05,Yonezawa07,Lee11} in one degree of freedom have been transported. Recent work has also demonstrated the capability of teleporting multiple degrees of freedom of a single photon \cite{Wang15}. However, to teleport quantum states of a real particle, for example, a single photon, one needs to consider not only the two-level states (polarization), but also those multilevel states. For example, the orbital angular momentum \cite{Dada11,Krenn14}, the temporal mode \cite{Martin17}, the frequency mode \cite{Kues17} and the spatial mode \cite{Schaeff15,Hu16,Hu20,Valencia20} of a single photon are all natural attributes of multilevel states, which are exploited as high-dimensional systems. However, to teleport high-dimensional quantum states is still a challenge for two reasons. One is the generation of high-quality high-dimensional entanglement feasible for quantum teleportation. There has been much work on high-dimensional entanglement generation \cite{Dada11,Krenn14,Martin17,Kues17,Schaeff15,Hu16,Hu20,Valencia20}, including attempts to observe interference between different high-dimensional entangled pairs \cite{Erhard18,Malik16}. Nevertheless, the interference visibility between different pairs is still quite low at $63.5\%$. The other concerns performing a deterministic high-dimensional Bell state measurement (HDBSM). Here, we use the spatial mode (path) to encode the three-dimensional states that has been demonstrated to extremely high fidelity \cite{Hu16} and use an auxiliary entangled photon pair to perform the HDBSM. We thereby overcome these obstacles and demonstrate the teleportation of a three-dimensional quantum state using the spatial mode of a single photon.

\begin{figure}[tbph]
\begin{center}
\includegraphics [width= 1\columnwidth]{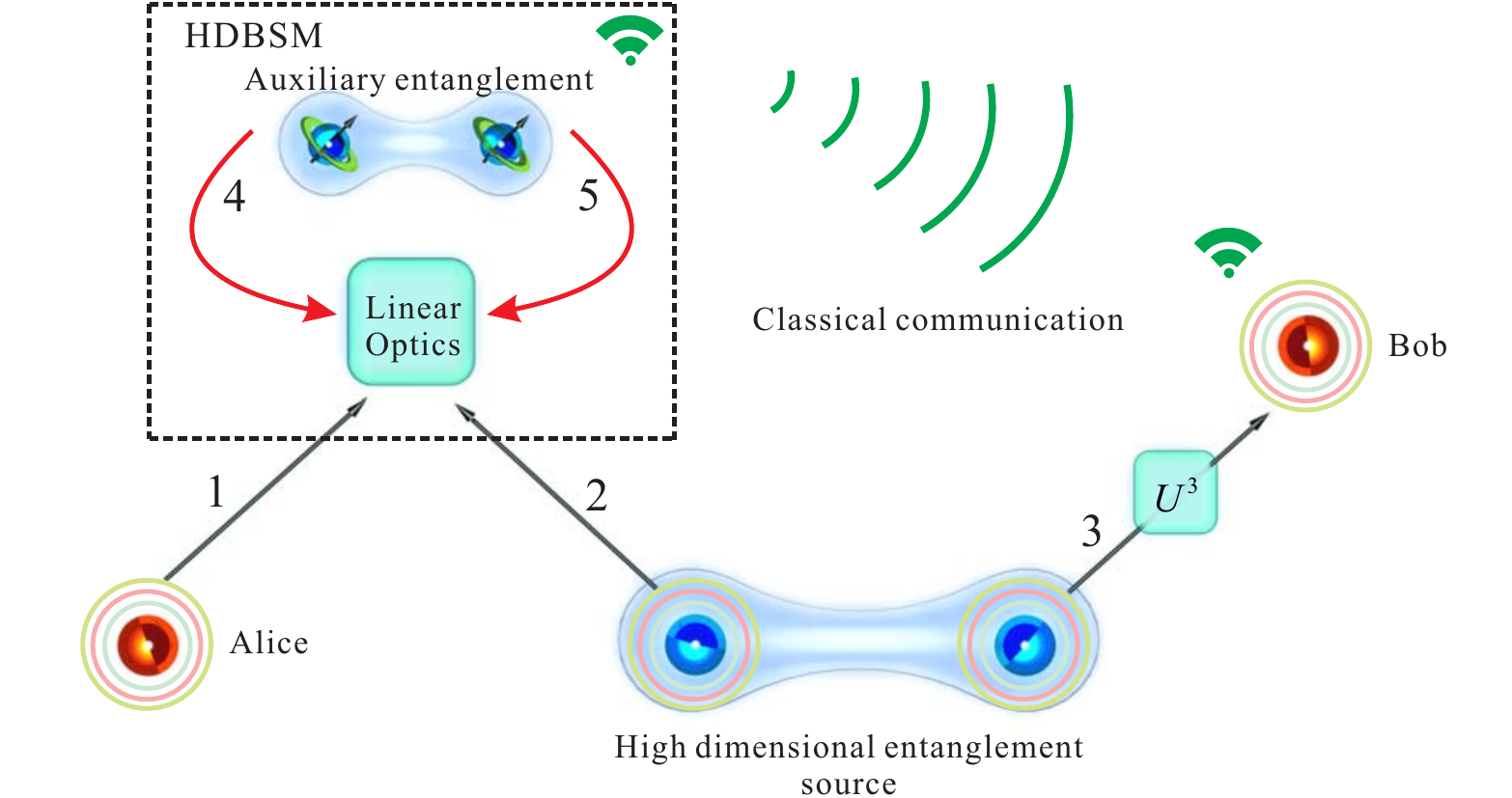}
\end{center}
\caption{\textbf{Scheme for quantum teleportation of the high-dimensional states of a single photon.} Alice wishes to teleport the high-dimensional quantum state of single photon 1 to Bob. Initially, Alice and Bob share a three-dimensional entangled photon pair 2--3. Then, Alice performs a high-dimensional Bell state measurement (HDBSM) assisted by another entangled photon pair 4--5 and sends the results to Bob through a classical channel. Finally, according to the results of HDBSM, Bob applies the appropriate three-dimensional Pauli operations on photon 3 to convert it into the original state of photon 1.}
\label{fig1}
\end{figure}

\begin{figure*}[tbph]
\begin{center}
\includegraphics [width= 1.9\columnwidth]{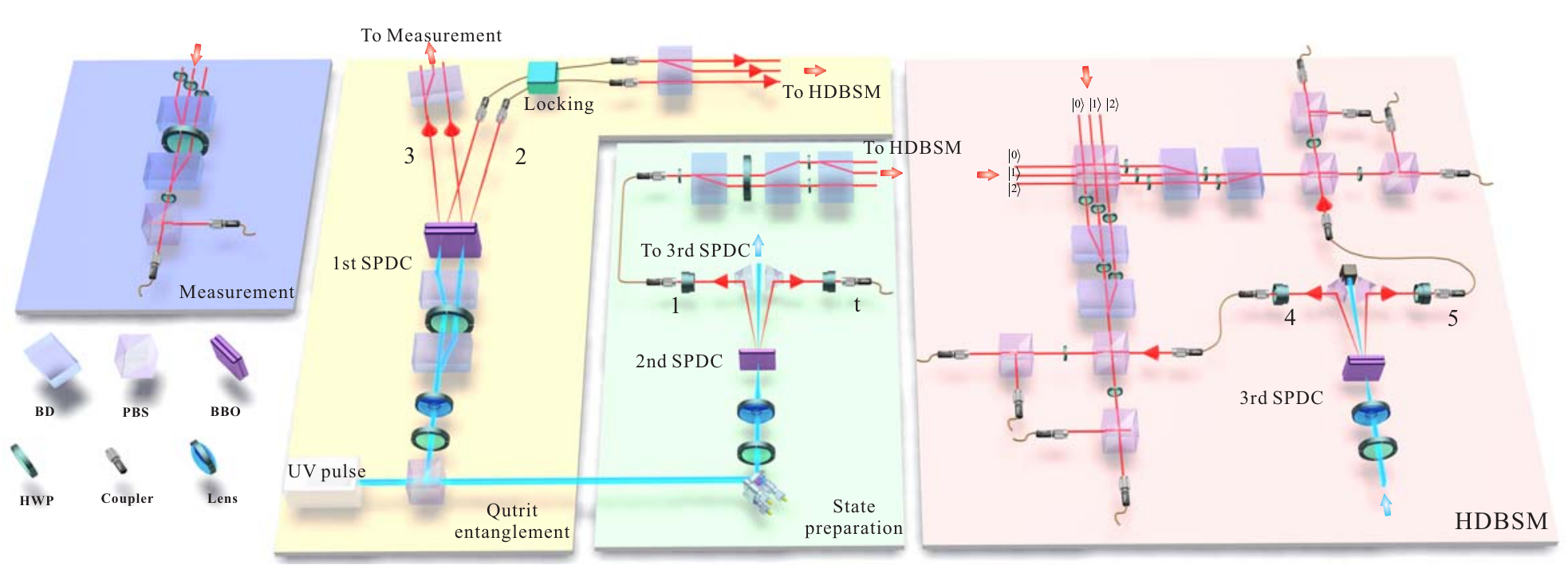}
\end{center}
\caption{\textbf{Experimental setup for teleporting a qutrit state of a single photon.} A pulsed ultraviolet (UV) laser is focused on three sets of $\beta$-barium borate (BBO) crystals and produces three photon pairs in 2--3, 1--t, and 4--5. The first pair, 2--3, is qutrit-qutrit entanglement in path DOF shared by Alice and Bob. The second pair, 1--t, photon 1 is initialized in various states ($|\varphi_{1}\rangle-|\varphi_{10}\rangle$) to be teleported, triggered by its twisted photon t. The third pair, 4--5, is a polarization-entangled state, used as an ancillary pair for performing a HDBSM on photons 1 and 2. BD-beam displacer, PBS-polarizing beam splitter, HWP-half wave plate.}
\label{fig2}
\end{figure*}

Suppose Alice wishes to teleport to Bob the quantum state of a single photon (photon 1, Fig.~\ref{fig1}), encoded in the path mode as

\begin{equation}
|\varphi\rangle=\alpha|0\rangle +\beta|1\rangle+\gamma|2\rangle,
\end{equation}
where $|0\rangle$, $|1\rangle$, and $|2\rangle$ denote the path degree of freedom (DOF). This DOF exists in an infinite dimensional space of the photonic system; here, we take only three dimensions as an example. The coefficients $\alpha$, $\beta$, and $\gamma$ are complex numbers satisfying $|\alpha|^{2}+|\beta|^{2}+|\gamma|^{2}=1$. Alice and Bob initially need to share a high-dimensional entangled photon pair (photons 2 and 3) in path

\begin{equation}
\begin{split}
|\xi\rangle_{23}=(|00\rangle_{23}+|11\rangle_{23}+|22\rangle_{23})/\sqrt{3}.
\end{split}
\end{equation}

Then, Alice performs a two-particle HDBSM on photons 1 and 2, which projects the two-photon state onto the basis of the nine orthogonal three-dimensional Bell states and discriminates one of them,
\begin{equation}
\begin{split}
&|\psi_{00}\rangle=(|00\rangle+|11\rangle+|22\rangle)/\sqrt{3},\\
&|\psi_{10}\rangle=(|00\rangle+e^{2\pi i/3}|11\rangle+e^{4\pi i/3}|22\rangle)/\sqrt{3},\\
&|\psi_{20}\rangle=(|00\rangle+e^{4\pi i/3}|11\rangle+e^{2\pi i/3}|22\rangle)/\sqrt{3},\\
&|\psi_{01}\rangle=(|01\rangle+|12\rangle+|20\rangle)/\sqrt{3},\\
&|\psi_{11}\rangle=(|01\rangle+e^{2\pi i/3}|12\rangle+e^{4\pi i/3}|20\rangle)/\sqrt{3},\\
&|\psi_{21}\rangle=(|01\rangle+e^{4\pi i/3}|12\rangle+e^{2\pi i/3}|20\rangle)/\sqrt{3},\\
&|\psi_{02}\rangle=(|02\rangle+|10\rangle+|21\rangle)/\sqrt{3},\\
&|\psi_{12}\rangle=(|02\rangle+e^{2\pi i/3}|10\rangle+e^{4\pi i/3}|21\rangle)/\sqrt{3},\\
&|\psi_{22}\rangle=(|02\rangle+e^{4\pi i/3}|10\rangle+e^{2\pi i/3}|21\rangle)/\sqrt{3}.\\
\end{split}
\end{equation}

After the HDBSM, photons 1 and 2 are projected onto the state $|\psi_{00}\rangle$ with a probability of 1/9, then photon 3 is projected onto state $|\varphi\rangle$. For instances where photons 1 and 2 are projected onto the other eight three-dimensional Bell states, Bob needs to perform a three-dimensional unitary operation on photon 3 to rotate the state of photon 3 to $|\varphi\rangle$ according to the measurement results of photons 1 and 2.

However, HDBSM is still a challenge with linear optics \cite{Calsamiglia02,Zhang19}. Although one can classify high-dimensional entangled states into several categories \cite{Hu182,Hill16}, one cannot identify any of them. The possible solution to this HDBSM is to introduce an auxiliary system. Here, we introduce a pair of assistant entangled photons to complete the HDBSM.

The Bell state measurement (BSM) of a two-dimensional polarized state is divided into two steps \cite{Liu16}. The four Bell states are first divided into two categories ($(|HH\rangle\pm|VV\rangle)/\sqrt{2}$ and $(|HV\rangle\pm|VH\rangle)/\sqrt{2}$) by a polarizing beam splitter (PBS) according to classical terms. Second, the two states are distinguished with different phases by projecting onto basis $|H\pm V\rangle/\sqrt{2}$. The structure of HDBSM in our system is similar to that of qubit polarized BSM. According to classical terms, nine three-dimensional Bell states are divided into three categories, then, the localized projection measurement is used to identify the three-dimensional Bell states.

\begin{figure}[tbph]
\begin{center}
\includegraphics [width=0.9\columnwidth]{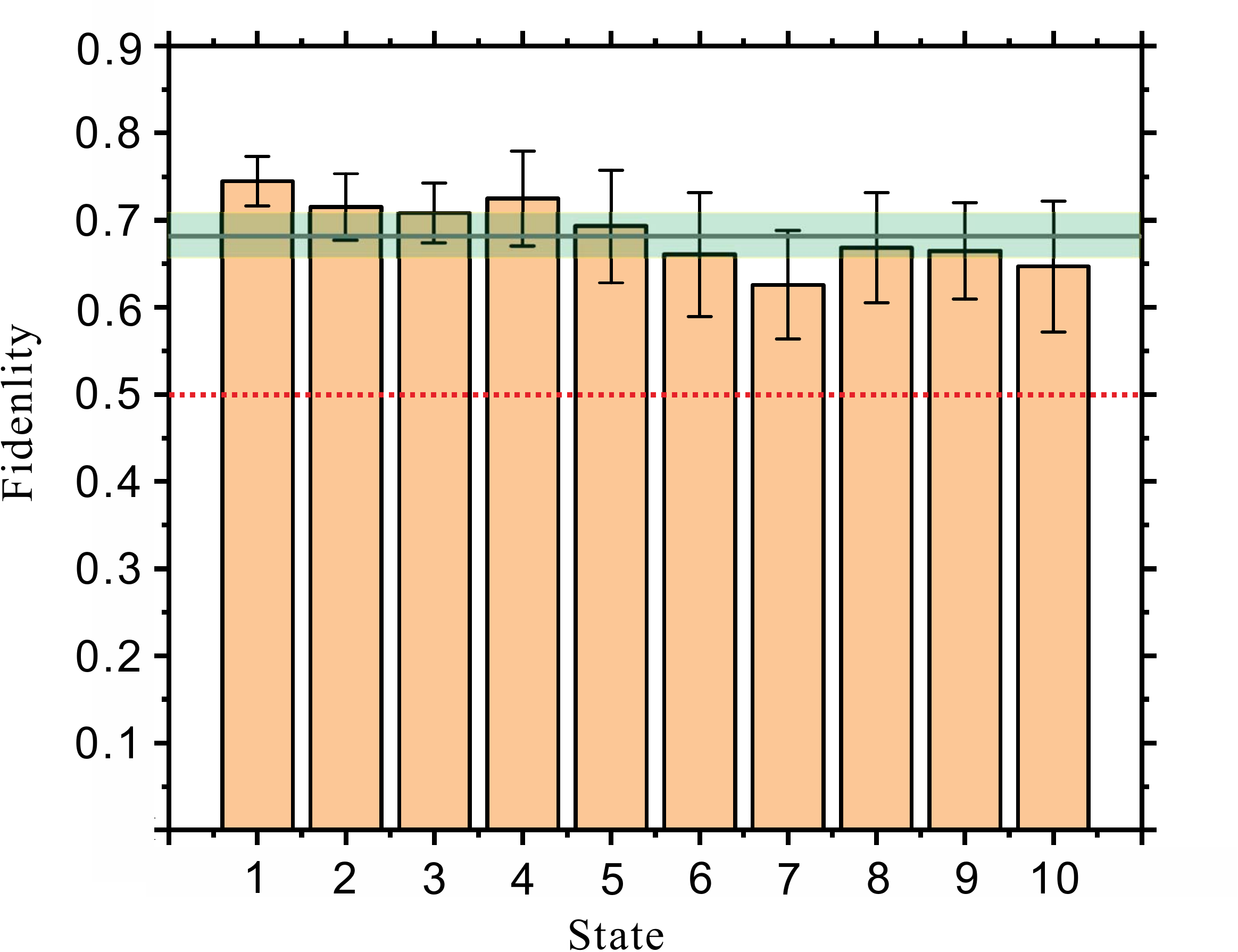}
\end{center}
\caption{\textbf{Experimental results for quantum teleportation of three-dimensional single photon states $|\varphi_{1}\rangle$-$|\varphi_{10}\rangle$.} We reconstruct the density matrix of the three-dimensional states by state tomography \cite{Thew2002}, and then obtain the fidelity of each state. Density matrix of the qutrits is reconstructed from a set of 9 measurements represented by
operators $u_{i}$(with i=1, 2,...9) and $u_{i}=|\Psi_{i}\rangle\langle\Psi_{i}|$. Kets $|\Psi_{i}\rangle$ are selected from the following setting $|0\rangle, |1\rangle, |2\rangle, (|0\rangle+|1\rangle)\sqrt{2}, (|0\rangle+i|1\rangle)\sqrt{2}, (|0\rangle+|2\rangle)\sqrt{2}, (|0\rangle+i|2\rangle)\sqrt{2}, (|1\rangle+|2\rangle)\sqrt{2}, (|1\rangle+i|2\rangle)\sqrt{2}$. Average fidelity($F=0.685\pm0.027$) of 10 states is significantly higher than that of non-classical quantum teleportation bound (0.5). Error bars are calculated from Poissonian counting statistics of the raw detection events.}
\label{fig3}
\end{figure}

Fig.~\ref{fig2} illustrates our linear optical scheme for teleporting the three-dimensional quantum states. The first step is to divide nine Bell states into three categories according to classical terms $|i,i\rangle$, $|i,i+1\rangle$, and $|i,i+2\rangle$, $i \in\{0,1,2\}$ under modulo-3 arithmetic. Photons 1 and 2 are sent to a PBS, which transmits horizontally polarized terms ($|H\rangle$) and reflects vertically polarized terms ($|V\rangle$). In the three-dimensional path state, we control the polarization of each path to satisfy ($|0\rangle\rightharpoonup|H\rangle$, $|1\rangle\rightharpoonup|V\rangle$, and $|2\rangle\rightharpoonup|H\rangle$). After the PBS, we post-select the event in which there is one and only one photon in each outport. For the nine classical terms of the three-dimensional Bell states ($|i,j\rangle$, $i,j\in\{0,1,2\})$, five of them are selected ($|i,j\rangle$ with $i+j$ even).

The second step is to use a local projection measurement to determine which Bell state is post-selected through $|\psi_{00}\rangle$, $|\psi_{10}\rangle$ and $|\psi_{20}\rangle$ (here, terms $|20\rangle$ and $|02\rangle$ are noise terms and are cancelled later). We can construct an arbitrary single qutrit basis (e.g., $(|0\rangle+|1\rangle+|2\rangle)/\sqrt{3}$) by half-wave plates (HWPs), beam displacers (BDs) and PBSs, so that we can determine whether the measured state is $|\psi_{00}\rangle$ by measuring this basis on both sides \cite{Hu16}.

To cancel the disturbance terms ($|20\rangle$ and $|02\rangle$), we introduce another entangled photon pair \cite{Supple}. Hence, we can distinguish at least one Bell state deterministically. For Bell states $|\psi_{10}\rangle$ and $|\psi_{20}\rangle$, we need to select different local projection measurements. Finally, we have teleported a three-dimensional quantum state with a success probability of 1/54. To increase the success probability, we use the non-maximally entangled state $(2|00\rangle+2|11\rangle+|22\rangle)/3$ to replace the maximally entangled state $|\xi\rangle_{23}$, and adjust the measurement base on HDBSM correspondingly, this increases the success probability of teleportation to 1/18 while does not affect the fidelity \cite{Supple}. This method of completing HDBSM in linear optical systems can be extended to higher dimensions and can be applied to different degrees of freedom such as orbital angular momentum (OAM). For d-dimensional systems, we only need $\lceil\log_{2}(d)\rceil-1$ pairs of auxiliary entangled photons \cite{Supple}.

The implementation of the HDBSM requires Hong-Ou-Mandel (HOM)-type interference between indistinguishable single photons with good temporal, spatial, and spectral overlap. We use a narrow band interference filter (3~nm) and a single-mode fiber to improve the visibility of HOM interference. For photon 3 and photon t, we use a broad band interference filter (8~nm) to increase the coincidence efficiency.

The verification of the teleportation results relies on the coincidence events of six photons. To suppress the statistical error, the data collection time is tens of hours. Hence, the stability of the whole system becomes a crucial aspect for the experiment. In our system, the HOM interference between the different photons is stable enough \cite{Zhang15,Hu19}, whereas the interference between different spatial modes after passing through the single-mode fibers is not. Here, we use a fiber phase locking system \cite{Supple} to maintain a phase-stable interferometer. The measured interference visibility remained above 0.98 in 45 hours \cite{Supple}.

\begin{figure*}[tbph]
\begin{center}
\includegraphics [width= 2\columnwidth]{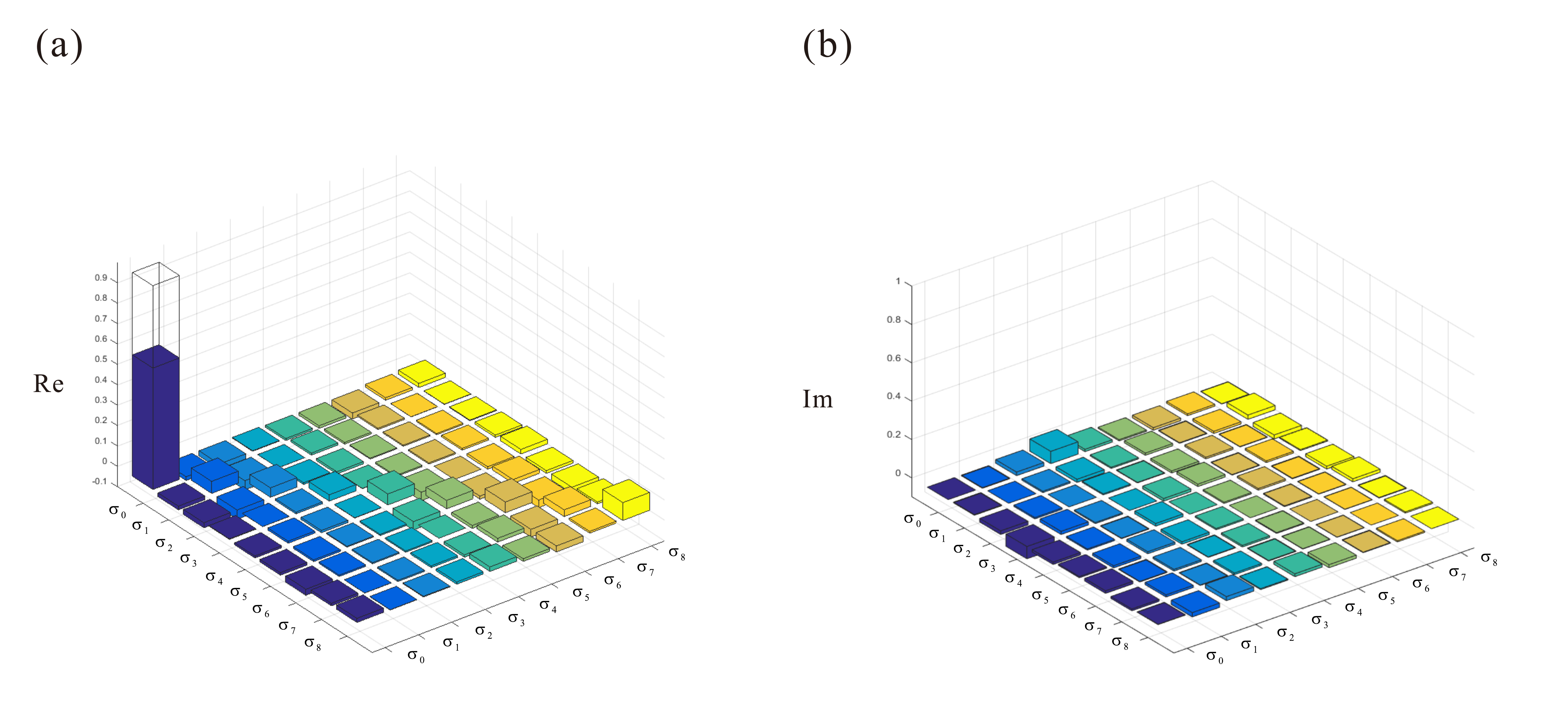}
\end{center}
\caption{\textbf{Quantum process tomography of three-dimensional quantum teleportation.} \textbf{a, b}, The real ($Re(\chi_{lk}$)) and imaginary ($Im(\chi_{lk})$) values of the
components of the reconstructed quantum process matrix, with l, k=0, 1, 2,..., 8. The results of the state tomography of the nine teleported states, $|\varphi_{1}\rangle-|\varphi_{9}\rangle$, are employed to reconstruct the process matrix of teleportation. The operator $\sigma_{0}-\sigma_{8}$ are Pauli matrices in three dimension. For the ideal case, the only non-zero component of the process matrix of quantum teleportation, $\chi_{ideal}$, is $(\chi_{ideal})_{00}=1$, which is indicated by transparent column. }
\label{fig4}
\end{figure*}

We prepared ten different initial states to be teleported: $|\varphi_{1}\rangle=|0\rangle$, $|\varphi_{2}\rangle=|1\rangle$, $|\varphi_{3}\rangle=|2\rangle$,
$|\varphi_{4}\rangle=(|0\rangle+|1\rangle)/\sqrt{2}$, $|\varphi_{5}\rangle=(|0\rangle+i|1\rangle)/\sqrt{2}$, $|\varphi_{6}\rangle=(|0\rangle+|2\rangle)/\sqrt{2}$, $|\varphi_{7}\rangle=(|0\rangle+i|2\rangle)/\sqrt{2}$, $|\varphi_{8}\rangle=(|1\rangle+|2\rangle)/\sqrt{2}$, $|\varphi_{9}\rangle=(|1\rangle+i|2\rangle)/\sqrt{2}$, and
$|\varphi_{10}\rangle=(|0\rangle+|1\rangle+|2\rangle)/\sqrt{3}$. The first nine states ($|\varphi_{1}\rangle$--$|\varphi_{9}\rangle$) constitute a complete orthogonal basis in three-dimensional space; the last state $|\varphi_{10}\rangle$ is a linear-dependent superposition of quantum states in this space. All these states are prepared by the BDs, QWPs, and HWPs.

To evaluate the performance of the high-dimensional teleportation, we reconstruct the density matrix of $|\varphi_{1}\rangle-|\varphi_{10}\rangle$ by state tomography. Conditioned on the detection of the trigger photon and the four-photon coincidence after the HDBSM, we registered the photon counts of teleported photon 1. As shown in Fig.~\ref{fig3}, the average fidelity of states $|\varphi_{1}\rangle-|\varphi_{10}\rangle$ is $F=0.685\pm0.027$, which is significantly higher than that of qutrit nonclassical teleportation bound ($>0.5$ \cite{Hayashi05,Brub99}).

All reported data are the raw data without background subtraction. The main sources of error include double pair emission, imperfect initial states, entanglement of photons 2--3 and 4--5, two-photon interference, and phase stabilization. We note that the teleportation fidelities of the states are affected differently by errors from the various sources. The fidelity of $|\varphi_{1}\rangle$-$|\varphi_{3}\rangle$ is higher than that of $|\varphi_{4}\rangle$-$|\varphi_{10}\rangle$. The reason is that imperfect interference does not affect the first three quantum states, but the latter.

The first nine states ($|\varphi_{1}\rangle-|\varphi_{9}\rangle$) are a set complete basis for three-dimensional tomography. The reconstructed density matrices of the teleported quantum states allow us to fully characterize the teleportation procedure by quantum process tomography \cite{Fiur2001}. We can completely describe the effect of teleportation on the input states $\rho_{ideal}$ by determining the process matrix $\chi$, defined by $\rho=\sum_{l, k=0}^{8} \chi_{l k} \sigma_{l} \rho_{ideal} \sigma_{k}$  where $\sigma_{0}-\sigma_{8}$ are the Pauli matrices for three dimension \cite{Thew2002}. The ideal process matrix of quantum teleportation $\chi_{ideal}$ has only one non-zero component $(\chi_{ideal})_{00}=1$, represents that the process of teleportation is perfect. Fig.~\ref{fig4} shows the real and imaginary components of $\chi$ for quantum teleportation based on our experimental results respectively. The process fidelity of our experiment was $f_{\text {process }}=\operatorname{Tr}\left(\chi_{ideal} \chi\right)=0.596\pm0.037$.

In general, to demonstrate that the three-dimensional teleportation is nonclassical using average fidelity, one need to measure 12 states from four mutually unbiased bases settings ($|\psi_{1}\rangle-|\psi_{12}\rangle$) \cite{Ivanovic1981,Supple}. In the three-dimensional case, the lower bound of average fidelity for nonclassical teleportation is 0.5. This condition can be converted to process fidelity \cite{Gilchrist05} and the lower bound of process fidelity is 1/3. In our experiment, the measured process fidelity is $f_{\text {process }}=0.596\pm0.037$, which is 7 standard deviations above the fidelity of 1/3, and proves that our teleportation is nonclassical.


For high-dimensional teleportation, it is not enough to only prove that teleportation is nonclassical. Genuine d-dimensional teleportation should be distinguished from the low dimensional case, excluding the hypothesis that the teleportation can be expressed in a smaller dimension. In our case, we need to exclude qubit and make sure that we have completed the genuine three-dimensional teleportation. In reference \cite{Luo2019}, two-dimensional states are transmitted, it is found that the maximum fidelity with state $(|0\rangle+|1\rangle+|2\rangle)/\sqrt{3}$ is 2/3. Therefore, quantum states with fidelity more than 2/3 are genuine three-dimensional states. However, this is a sufficient but not a necessary condition. For some states (like $\sqrt{1/8}|0\rangle+\sqrt{1/8}|1\rangle-\sqrt{3/4}|2\rangle$), the fidelity can not reach 2/3, but they are still genuine three-dimensional coherent superposition states. If someone who can transmit all qubit states is unable to simulate a teleportation, then it is reasonable to say that the teleportation performed is genuine three dimensional. We assume that the qubit state ($\rho_{qubit}=(P_{1}\rho_{01}+P_{2}\rho_{02}+P_{3}\rho_{12}$) is incoherent at different subspaces ($\{|0\rangle,|1\rangle\}, \{|0\rangle,|2\rangle\}, \{|1\rangle,|2\rangle\}$). If we cannot use this qubit state to simulate the state after teleportation, then we prove that our teleportation state is in the state of three levels of coherent superposition. First, we derive a nonlinear criterion \cite{Supple}, which is more powerful than the fidelity criterion. This criterion can be used to determine states like $\sqrt{1/8}|0\rangle+\sqrt{1/8}|1\rangle-\sqrt{3/4}|2\rangle$ are genuine three-dimensional states. Of course, this criterion is still not a necessary and sufficient condition. We define the robustness ($\mu$) \cite{Cavalcanti2017} of the genuine three-dimensional state. The optimal solution $\mu^{*}$ of this problem gives the minimum amount of "white noise" that has to be added to the qutrit state such that the mixture can be simulated by qubit states \cite{Supple}. If $\mu>0$, we can certify that this state is a genuine three-dimensional state. On the contrary, if $\mu=0$, the state is not a genuine three-dimensional state. We choose 400 maximum coherent superposition states ($1/\sqrt{3}(|0\rangle+e^{i\varphi_{1}}|1\rangle+e^{i\varphi_{2}}|2\rangle)$,where $\varphi_{1},\varphi_{2}$ are 20 phases at equal interval in $[0,\pi)$) as the input states, and then through the evolution of $\chi$ matrix. After the semidefinite program (SDP), we find that 149 states can be simulated by qubit, while 251 states cannot be simulated. The average $\mu$ of these states that can not be simulated is $\mu=0.111\pm0.034>0$. This means that within three standard deviation ranges, using qubit states cannot simulate our teleportation process. This result can prove that our teleportation is beyond qubit. All the errors are obtained by raw data through the Monte Carlo method, in which all the generated data have the same Poissonian error as the raw data.

In summary, we have reported the quantum teleportation of high-dimensional quantum states of a single quantum particle, demonstrating the capability to control coherently and teleport simultaneously a high-dimensional state of a single object. The generation of high-quality high-dimensional multi-photon state will stimulate the research on high-dimensional quantum information tasks, and entanglement-assisted methods for HDBSM are feasible for other high-dimensional quantum information tasks.

\begin{acknowledgments}
We thank Che-Ming Li for the valuable information. This work was supported by the National Key Research and Development Program of China (No.\ 2017YFA0304100, No. 2016YFA0301300, and No. 2016YFA0301700), NSFC (No. 11774335, No. 11734015, No. 11874345, No. 11821404, No. 11805196, and No. 11904357), the Key Research Program of Frontier Sciences, CAS (No.\ QYZDY-SSW-SLH003), Science Foundation of the CAS (ZDRW-XH-2019-1), the Fundamental Research Funds for the Central Universities, Science and Technological Fund of Anhui Province for Outstanding Youth (2008085J02), Anhui Initiative in Quantum Information Technologies (Nos.\ AHY020100, AHY060300) and Swiss national science foundation (Starting grant DIAQ, NCCR-QSIT).\\
\end{acknowledgments}

\clearpage
\setcounter{table}{0}
\renewcommand{\thetable}{S\arabic{table}}
\setcounter{figure}{0}
\renewcommand{\thefigure}{S\arabic{figure}}
\setcounter{equation}{0}
\renewcommand{\theequation}{S\arabic{equation}}

\onecolumngrid
\textbf{SUPPLEMENTARY INFORMATION}
\bigskip

\noindent\textbf{Protocol for teleporting three-dimensional quantum states.} The combined state of photons 1, 2, and 3 are rewritten in terms of basis states formed from nine orthogonal three-dimensional Bell states as follows:
\begin{equation}
\begin{split}
|\xi\rangle_{23}|\varphi\rangle_{1}=
&\frac{1}{\sqrt{3}}(|00\rangle_{23}+|11\rangle_{23}+|22\rangle_{23})(\alpha|0\rangle_1+\beta|1\rangle_1+\gamma|2\rangle_1)\\
=&\frac{1}{3}(|\psi_{00}\rangle_{12}\otimes U_{00}^{\dag}|\varphi\rangle_{3}+|\psi_{10}\rangle_{12}\otimes U_{10}^{\dag}|\varphi\rangle_{3}+|\psi_{20}\rangle_{12}\otimes U_{20}^{\dag}|\varphi\rangle_{3}  \\
&+|\psi_{01}\rangle_{12}\otimes U_{01}^{\dag}|\varphi\rangle_{3}+|\psi_{11}\rangle_{12}\otimes U_{11}^{\dag}|\varphi\rangle_{3}+|\psi_{21}\rangle_{12}\otimes U_{21}^{\dag}|\varphi\rangle_{3}  \\
&+|\psi_{02}\rangle_{12}\otimes U_{02}^{\dag}|\varphi\rangle_{3}+|\psi_{12}\rangle_{12}\otimes U_{12}^{\dag}|\varphi\rangle_{3}+|\psi_{22}\rangle_{12}\otimes U_{22}^{\dag}|\varphi\rangle_{3}).
\end{split}
\end{equation}

The three-dimensional Pauli matrices are:
\begin{equation}
\begin{split}
&U_{00}=\left(
\begin{array}{ccc}
1 & 0 & 0\\
0 & 1 & 0\\
0 & 0 & 1\\
\end{array}
\right),\
U_{01}=\left(
\begin{array}{ccc}
0 & 0 & 1\\
1 & 0 & 0\\
0 & 1 & 0\\
\end{array}
\right), \
U_{02}=\left(
\begin{array}{ccc}
0 & 1 & 0\\
0 & 0 & 1\\
1 & 0 & 0\\
\end{array}
\right), \\
&U_{10}=\left(
\begin{array}{ccc}
1 & 0 & 0\\
0 & e^{\frac{2\pi i}{3}} & 0\\
0 & 0 & e^{\frac{4\pi i}{3}}\\
\end{array}
\right),\
U_{11}=\left(
\begin{array}{ccc}
0 & 0 & e^{\frac{4\pi i}{3}}\\
1 & 0 & 0\\
0 & e^{\frac{2\pi i}{3}} & 0\\
\end{array}
\right), \
U_{12}=\left(
\begin{array}{ccc}
0 & e^{\frac{2\pi i}{3}} & 0\\
0 & 0 & e^{\frac{4\pi i}{3}}\\
1 & 0 & 0\\
\end{array}
\right), \\
&U_{20}=\left(
\begin{array}{ccc}
1 & 0 & 0\\
0 & e^{\frac{4\pi i}{3}} & 0\\
0 & 0 & e^{\frac{2\pi i}{3}}\\
\end{array}
\right),\
U_{21}=\left(
\begin{array}{ccc}
0 & 0 & e^{\frac{2\pi i}{3}}\\
1 & 0 & 0\\
0 & e^{\frac{4\pi i}{3}} & 0\\
\end{array}
\right), \
U_{22}=\left(
\begin{array}{ccc}
0 & e^{\frac{4\pi i}{3}} & 0\\
0 & 0 & e^{\frac{2\pi i}{3}}\\
1 & 0 & 0\\
\end{array}
\right). \\
\end{split}
\end{equation}


By performing HDBSMs on photons 1 and 2, Alice determines which Bell state she has, for example, $|\psi_{00}\rangle$. When she tells Bob the measurement result, Bob applies an appropriate three-dimensional unitary operation on photon 3 and recovers the quantum state of photon 1.\\

\noindent\textbf{Generating three photon pairs.} The ultraviolet pulse laser from the mode-locked Ti:sapphire laser is split into two beams by a PBS. One beam (with 170~mW power) is used to generate a qutrit entanglement source. The other beam (with 170~mW power) successively passes through two "beamlike" sources to generate the two-photon pairs in type-II spontaneous parametric down conversion (SPDC).

\begin{figure}[tbph]
\begin{center}
\includegraphics [width= 0.5\columnwidth]{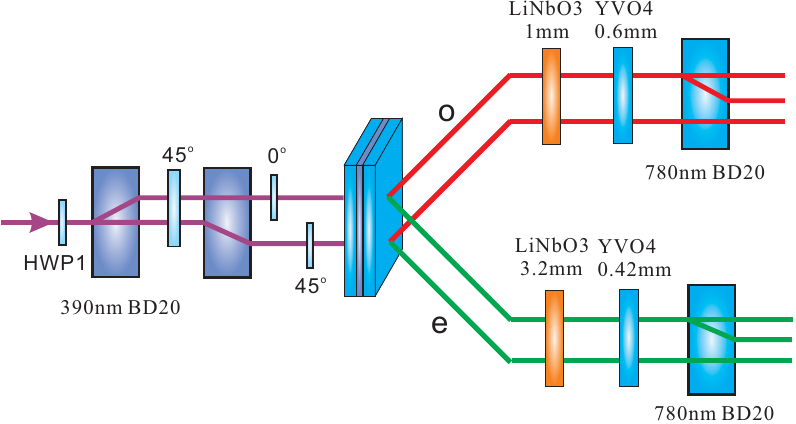}
\end{center}
\caption{\textbf{Experimental device for three-dimensional entanglement preparation.} The pump beam is split into two beams by two 390-nm beam displacers (BD20s) and HWPs and focused on a type-II beamlike cut BBO crystal at two different spots to generate photon pairs via SPDC. The two BD20 axes are opposite to keep the upper and lower light paths equal. The upper path generates polarization entanglement $(|HH\rangle+|VV\rangle)/\sqrt{2}$ and therefore needs time and space compensation. The energy splitting ratio of upper and lower pumping beams is 4:1. Finally, the $|H\rangle$ and $|V\rangle$ beams are divided into parallel beams by BD20 so we can generate the state $(2|00\rangle+2|11\rangle+|22\rangle)/3$.}
\vspace{-0.5cm}
\label{fig5}
\end{figure}

The qutrit entanglement source is different from the previous three-dimensional entanglement source we have prepared \cite{Hu18}. The source here is pumped by an ultrashort pulsed laser using a type II BBO crystal with a beam-like sandwich structure \cite{Zhang15}. Hence, the whole optical path must be compensated strictly in the temporal domain and spatial domain (see Fig. \ref{fig5}). Finally, we used a BD20 to divide $|H\rangle$ and $|V\rangle$ polarization into two parallel beams to form a three-dimensional entanglement of paths. In this experiment, we ignored the $|VV\rangle$ photon pairs generated by the lower path. The interference visibility of any two-dimensional subspace ($(|0\rangle\pm|1\rangle)/\sqrt{2}$, $(|0\rangle\pm|2\rangle)/\sqrt{2}$, $(|1\rangle\pm|2\rangle)/\sqrt{2}$) is greater than 0.98. To maximize the probability of successful teleportation, the entangled state we actually prepared is $(2|00\rangle+2|11\rangle+|22\rangle)/3$.

The teleported high-dimensional quantum state is triggered by the detection of photon t. HWPs, QWPs, and BD20 are used to prepare our teleported quantum states. The auxiliary entangled pairs are generated by the standard technology \cite{Zhang15}.

All of the down-converted photons have central wavelengths of 780~nm. The first photon pair (2--3) is qutrit entangled and has a count rate of $4\times10^{4}$~Hz and a state fidelity of 0.98. The second photon pair (1--t) is initially prepared in a path degree of freedom with a coincidence count rate of $7.2\times10^{4}$~Hz. The third photon pair (4--5) is created in the polarization entangled state with a count rate of $4.6\times10^{4}$~Hz and a fidelity of 0.98. We estimate the mean numbers of photon pairs generated per pulse as 0.013, 0.015, and 0.031 for the first, second, and third pairs, respectively.\\

\begin{figure}[tbph]
\begin{center}
\includegraphics [width= 0.5\columnwidth]{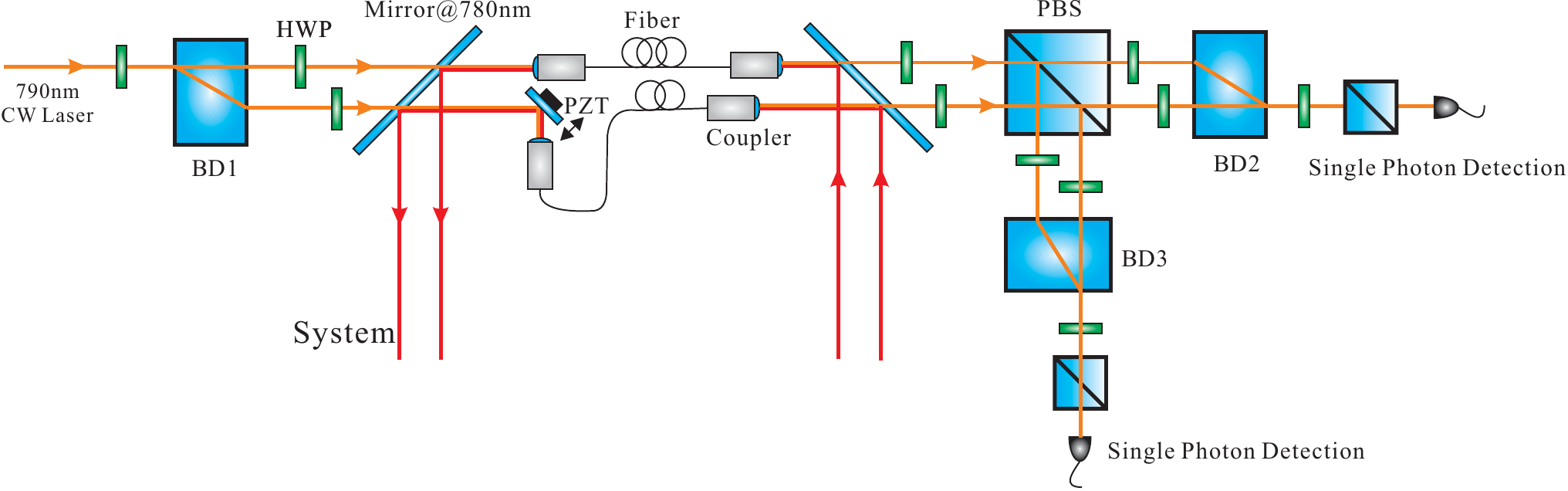}
\end{center}
\caption{\textbf{Optical fiber phase locking system.} To reduce the influence of the reference light (orange solid line) on the signal light (red solid line) of the system, their propagation directions oppose each other. The light from both are weak. The whole system consists of two MZ interferometers, BD1--BD2 and BD1--BD3. The first interferometer is used to provide the phase-locked signal, and the other interferometer is used to monitor the phase-locked interference visibility. }
\vspace{-0.5cm}
\label{fig6}
\end{figure}

\noindent\textbf{Fiber phase locking.} In our experimental scheme, we filtered each light source into a single mode, so we need to ensure the phase stability between the two optical fibers in high-dimensional entanglement. We used active feedback to lock the phase of the optical fiber. To reduce the disturbance of the quantum state by optical-fiber phase locking, the reference light (790~nm CW light) is opposite to the system light. The reference light passes through BD1 and is divided into two beams (see Fig. \ref{fig6}), which are coupled to single-mode fibers using mirrors. The light in the lower path needs to be reflected by the mirror of a piezoelectric ceramic material (PZT), which is used to change the length of the optical path to stabilize the phase.

\begin{figure}[tbph]
\begin{center}
\includegraphics [width= 0.5\columnwidth]{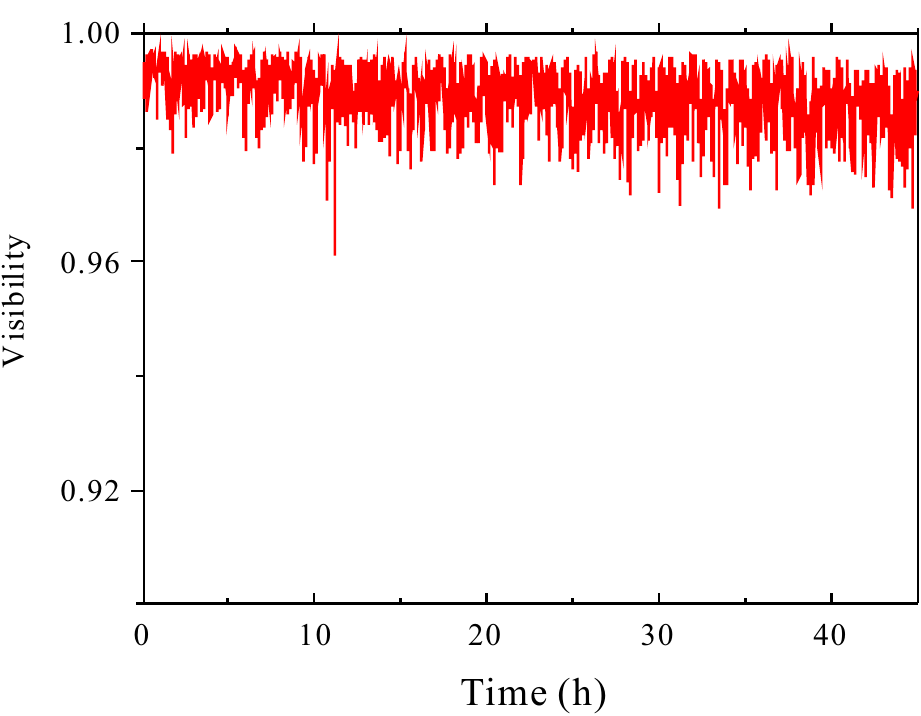}
\end{center}
\caption{\textbf{Fiber lock-in phase stabilization.} Within 45 hours, the average interference visibility is above 0.98. The fluctuation of interference visibility is mainly affected by ambient temperature and power stability of reference light.}
\vspace{-0.5cm}
\label{fig7}
\end{figure}

The reference light is then emitted from the other end of the optical fiber, and the two Mach--Zehner (MZ) interferometers are constructed by pairs BD1--BD2 and BD1--BD3. The phase between BD is stable, and all the phase changes are due to the instability of the two optical fibers. Hence, we only need to use the PZT to adjust the position of the mirror according to the signal of the MZ interferometer composed of BD1 and BD2 to lock the phase of the optical fiber. The other interferometer consisting of BD1 and BD3 is used to monitor the interference visibility of the optical fiber system. In this system, weak light is used as reference light, and a single photon detector is used for detection. We monitored the interference visibility of our system for 45 hours, and the results show that the average interference visibility remained around 0.98 (see Fig. \ref{fig7}).\\

\noindent\textbf{Analysis of noise in the experiment.}
Some error sources affect all the teleported states. First, the imperfectly entangled photon pairs. The first photon pair (2--3) is qutrit entangled and has a count rate of $4\times10^{4}$~Hz and a state fidelity of $\sim0.98$. The second photon pair (1--t) is initially prepared in a path DOF with a coincidence count rate of $7.2\times10^{4}$~Hz. The third photon pair (4--5) is created in the polarization-entangled state with a count rate of $4.6\times10^{4}$~Hz and fidelity of $\sim0.98$. The coincidence efficiencies of the three sources are $\eta_{1}\sim20\%$, $\eta_{2}\sim26\%$, and $\eta_{3}\sim14\%$. Second, all kinds of polarization setups are not ideal. The fidelity effect of these setups is $\sim2\%$. Third, the double pair emission, $\sim20\%$ noise is introduced.

\begin{figure}[tbph]
\begin{center}
\includegraphics [width= 0.5\columnwidth]{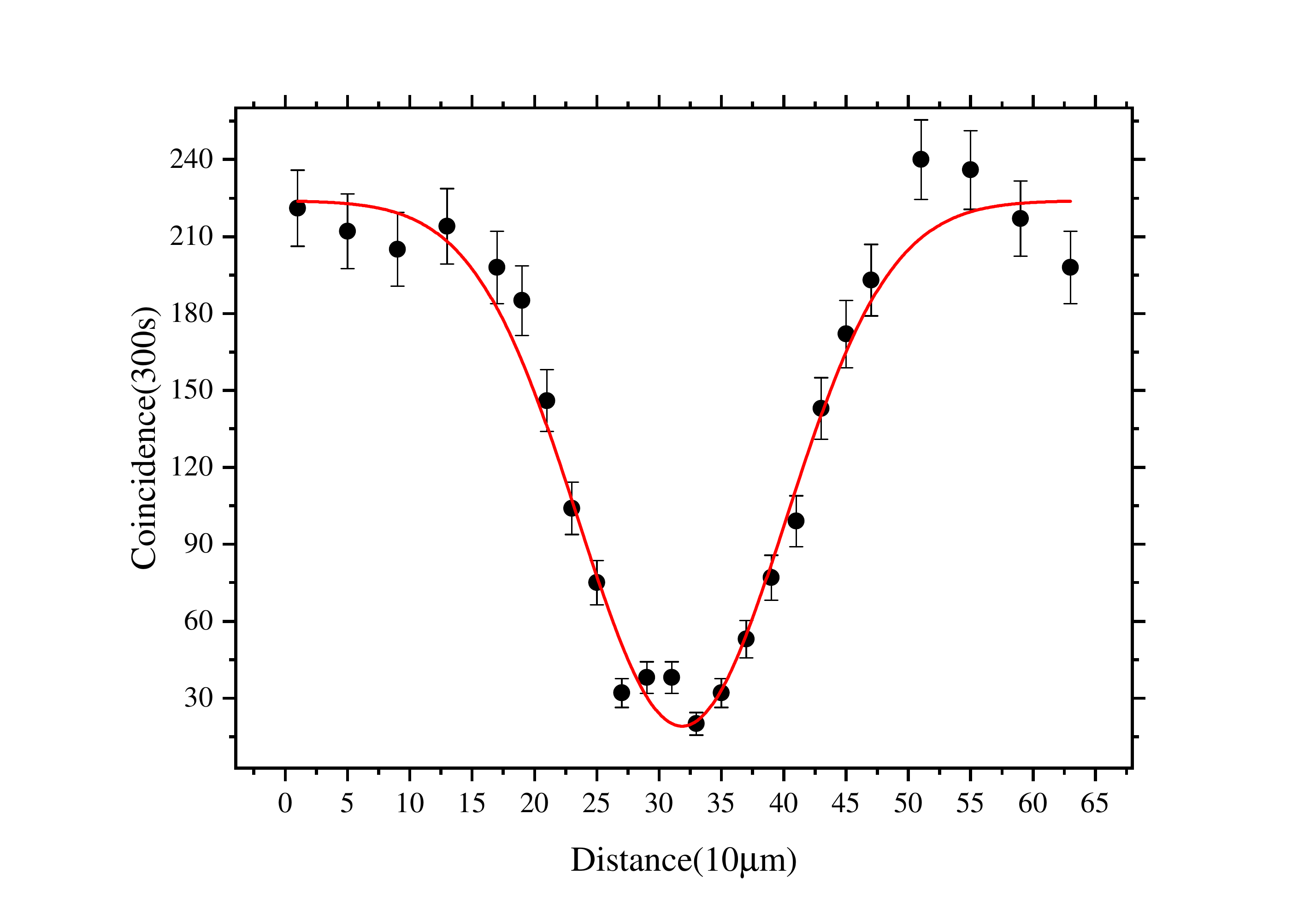}
\end{center}
\caption{\textbf{3nm-8nm and 3nm-8nm HOM interference.} HOM interference between two independent sources. This is the interference between photon pair 1-t and photon pair 2-3. Here, the full width at half maximum (FWHM) of the interference filters for photon 1 and photon 2 is 3~nm, and is 8~nm for photon 3 and photon t. The interference visibility is $V=0.918$.}
\label{fig8}
\vspace{-0.5cm}
\end{figure}

\begin{figure}[tbph]
\begin{center}
\includegraphics [width= 0.5\columnwidth]{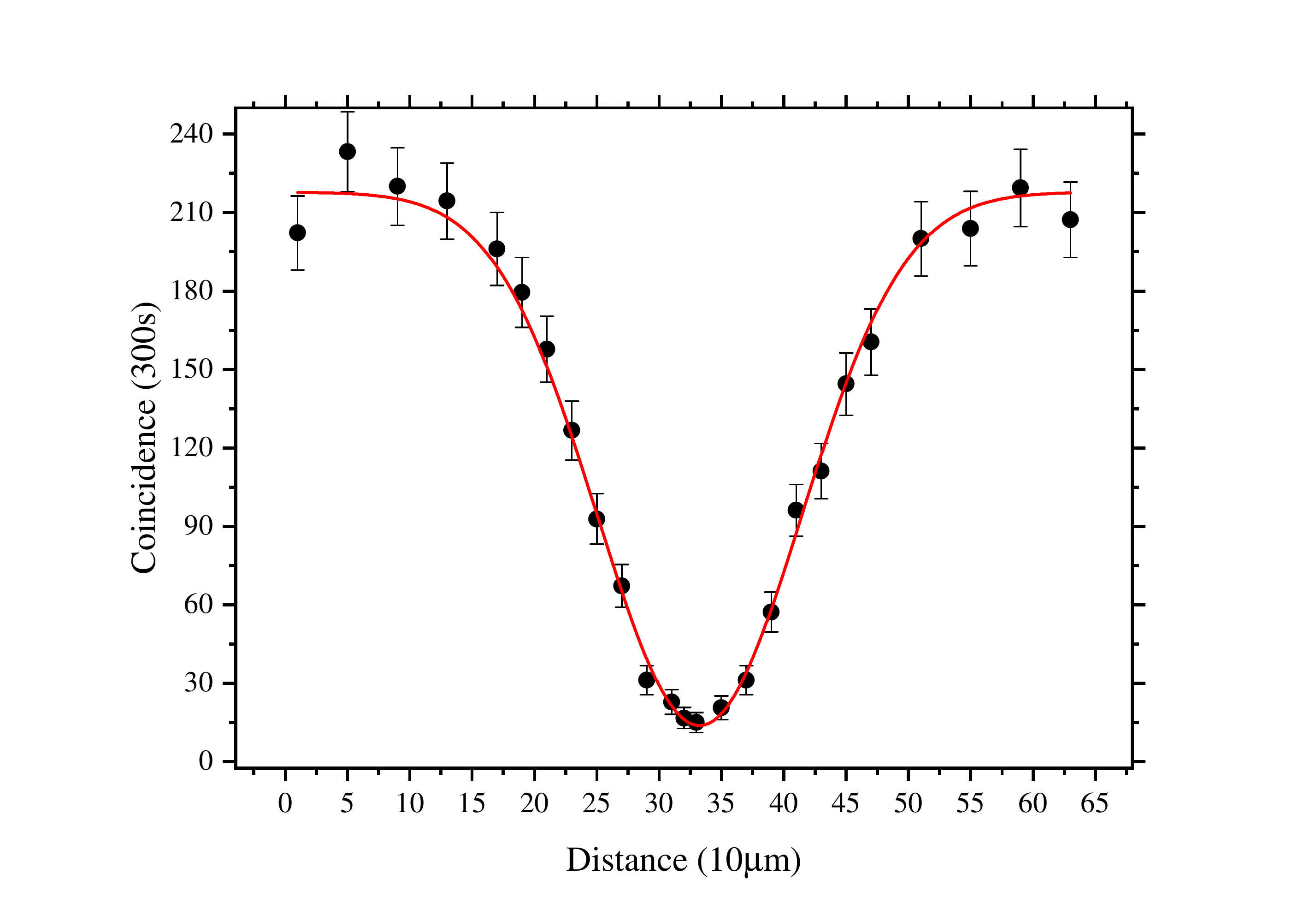}
\end{center}
\caption{\textbf{3nm-3nm and 3nm-3nm HOM interference.} HOM interference between two independent sources. In order to improve the interference visibility of HOM, we use 3~nm (FWHM) interference filters for all the photons. The interference visibility is $V=0.935$.}
\label{fig9}
\vspace{-0.5cm}
\end{figure}

Some error sources can have different effects on different teleported states. The visibilities of HOM interference of two photons is $\sim0.927$ (Fig. \ref{fig8} and Fig. \ref{fig9}). The imperfect two-photon interference at the PBS degrades the teleportation fidelities for the states $|\varphi_{4}\rangle-|\varphi_{10}\rangle$ by $ \sim5\%$, the noise from the locking system is $\sim2\%$. These two noises do not affect $|\varphi_{1}\rangle-|\varphi_{3}\rangle$, so the fidelity of $|\varphi_{1}\rangle-|\varphi_{3}\rangle$ are significantly higher than that of $|\varphi_{4}\rangle-|\varphi_{10}\rangle$. \\

\noindent\textbf{Three-dimensional BSM.} In this section, we introduce the 3-HDBSM in detail. There are nine three-dimensional Bell states $|\psi_{ij}\rangle$:
\begin{equation}
\begin{split}
&|\psi_{00}\rangle=(|00\rangle+|11\rangle+|22\rangle)/\sqrt{3},\\
&|\psi_{10}\rangle=(|00\rangle+e^{2\pi i/3}|11\rangle+e^{4\pi i/3}|22\rangle)/\sqrt{3},\\
&|\psi_{20}\rangle=(|00\rangle+e^{4\pi i/3}|11\rangle+e^{2\pi i/3}|22\rangle)/\sqrt{3},\\
&|\psi_{01}\rangle=(|01\rangle+|12\rangle+|20\rangle)/\sqrt{3},\\
&|\psi_{11}\rangle=(|01\rangle+e^{2\pi i/3}|12\rangle+e^{4\pi i/3}|20\rangle)/\sqrt{3},\\
&|\psi_{21}\rangle=(|01\rangle+e^{4\pi i/3}|12\rangle+e^{2\pi i/3}|20\rangle)/\sqrt{3},\\
&|\psi_{02}\rangle=(|02\rangle+|10\rangle+|21\rangle)/\sqrt{3},\\
&|\psi_{12}\rangle=(|02\rangle+e^{2\pi i/3}|10\rangle+e^{4\pi i/3}|21\rangle)/\sqrt{3},\\
&|\psi_{22}\rangle=(|02\rangle+e^{4\pi i/3}|10\rangle+e^{2\pi i/3}|21\rangle)/\sqrt{3}.\\
\end{split}
\end{equation}

\begin{figure}[tbph]
\begin{center}
\includegraphics [width= 0.5\columnwidth]{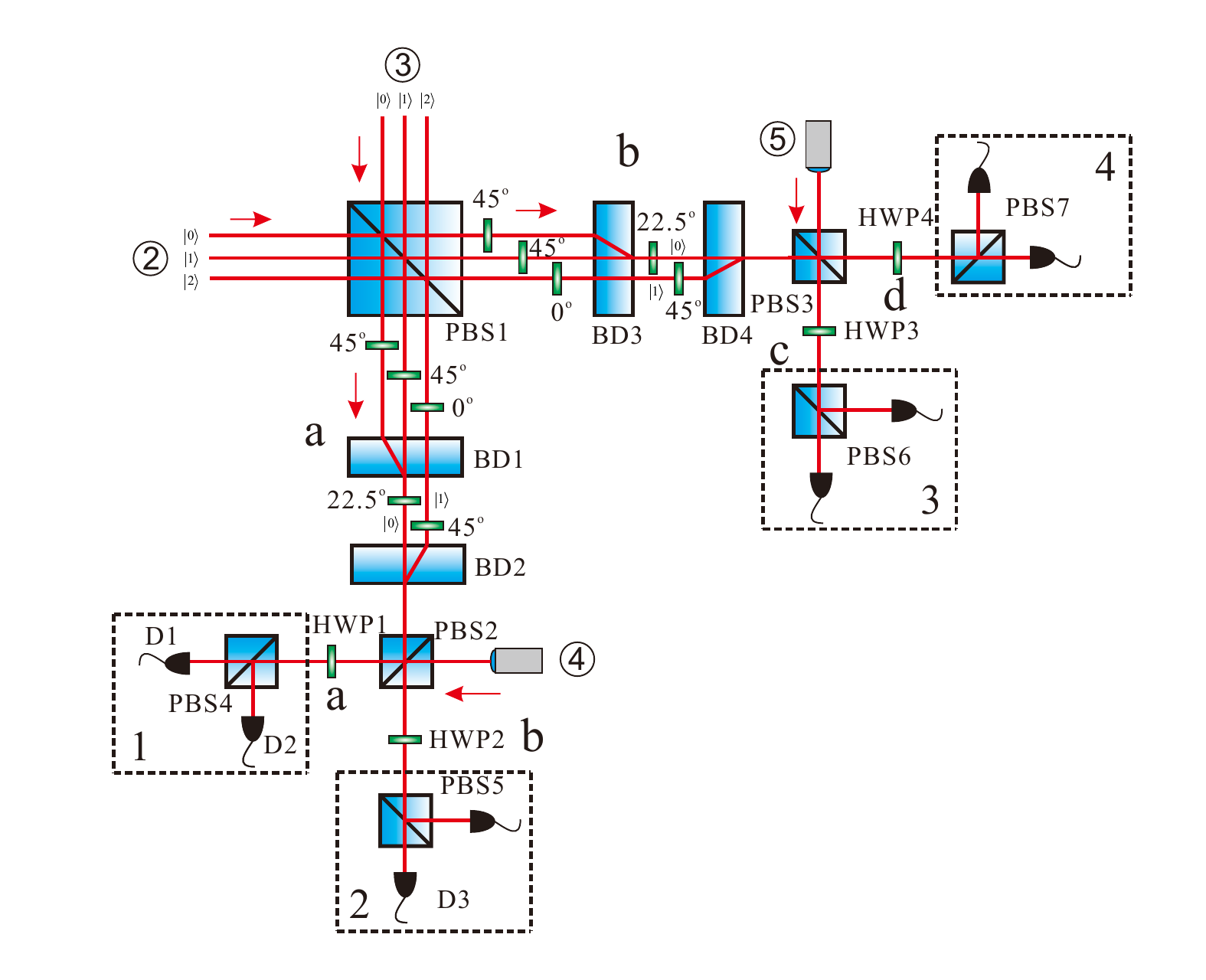}
\end{center}
\caption{\textbf{Three dimension BSM scheme.} When two three-dimensional photons are incident on the PBS, we only retain terms $|00\rangle$, $|11\rangle$, $|22\rangle$, $|02\rangle$, and $|20\rangle$ where there is only one photon on each outport. Noise terms $|02\rangle$, $|20\rangle$ are filtered out by the auxiliary entanglement.}
\label{fig10}
\vspace{-0.5cm}
\end{figure}

Similar to the measurements of the two-dimensional Bell states, the first step is to classify Bell states according to classical terms. For three-dimensional Bell states, we need to classify them into three categories: $\{|\psi_{i,0}\rangle; i=0,1,2\}$, which contain $|00\rangle$, $|11\rangle$, and $|22\rangle$; $\{|\psi_{i,1}\rangle; i=0,1,2\}$, which contain $|01\rangle$, $|12\rangle$ and $|20\rangle$; and $\{|\psi_{i,2}\rangle; i=0,1,2\}$, which contain $|02\rangle$, $|10\rangle$, and $|21\rangle$. We convert the path information to polarization-path hybrid encoded states (see Fig. \ref{fig10}), such that $|0\rangle\rightharpoonup|H\rangle_0$, $|1\rangle\rightharpoonup|V\rangle_1$, $|2\rangle\rightharpoonup|H\rangle_2$.

After the two photons enter the PBS1, only $|00\rangle$, $|11\rangle$, $|22\rangle$, $|02\rangle$, and $|20\rangle$ can have one photon on each outport of PBS1; the other classical terms $|01\rangle$, $|12\rangle$, $|10\rangle$, and $|21\rangle$ have two photons on only one side. If we only collect the events for which two photons exit on the two sides of PBS1, then the latter four classical terms are ignored. Here we only consider three terms $|00\rangle$, $|11\rangle$, and $|22\rangle$ as signals (there are still two noise terms $|02\rangle$ and $|20\rangle$, which can be cancelled by introducing an auxiliary entanglement to be explained later), and can distinguish the three Bell states $|\psi_{00}\rangle$, $|\psi_{10}\rangle$, and $|\psi_{20}\rangle$ through local projection measurements from BD3--BD6, PBSs, and HWPs. For example, if we construct a measurement basis of $(|0\rangle+|1\rangle+|2\rangle)/\sqrt{3}$ on both sides, then we can identify the quantum state of $(|00\rangle+|11\rangle+|22\rangle)/\sqrt{3}$.

We now explain the removal of the two noise terms $|02\rangle$, $|20\rangle$. In our scheme, we use an auxiliary entangled photon pair 4--5 to cancel the noise terms. If we consider the optical path between BD2 and BD4 as a process in changing the quantum state, we find that after BD3 and BD4, classical terms $|00\rangle$, $|11\rangle$, and $|22\rangle$ convert to $|HH\rangle$ or $|VV\rangle$, whereas noise terms $|02\rangle$ and $|20\rangle$ convert to $|HV\rangle$ and $|VH\rangle$. If we only collect those events for which there is one and only one photon at each outport of PBS3 and PBS4, then the two noise terms can be cancelled.

In three dimension BSM, we identify only one of the nine Bell states deterministically, and we use only one measurement basis for the two sides, so the probability of success is only $1/9\times1/3=1/27$. In addition, the probability of success for auxiliary entangled photons is $1/2$. The overall efficiency of the three dimension BSM combining all steps is $1/9\times1/3\times1/2=1/54$.

To improve the success probability, we do not use maximum entanglement $(|00\rangle+|11\rangle+|22\rangle)/\sqrt{3}$, but use $(2|00\rangle+2|11\rangle+|22\rangle)/3$ as the entangled channel for teleportation. Correspondingly, $(|H\rangle\pm|V\rangle)/\sqrt{2}$ is used for the projection measurement on four photons in the HDBSM. In this way, we improve the success probability of teleportation to 1/18.

Suppose we prepare a three-dimensional entangled state $|\xi_{23}\rangle=(2|H_{0}\rangle|H_{0}\rangle+2|V_{1}\rangle|V_{1}\rangle+|H_{2}\rangle|H_{2}\rangle)/3$ on photon 2--3. The teleported state (photon 1) and the trigger state (photon t) is prepared on $|\varphi\rangle_{1}\otimes|H\rangle_{t}=(\alpha|H_{0}\rangle +\beta|V_{1}\rangle+\gamma|H_{2}\rangle)\otimes|H\rangle_{t}$. Therefore, the quantum state of photon 1, 2, 3, and t is

\begin{equation}
\begin{split}
|\varphi\rangle_{1}\otimes|\xi\rangle_{23}\otimes|H\rangle_{t}
=&(\alpha|H_{0}\rangle+\beta|V_{1}\rangle+\gamma|H_{2}\rangle)\otimes(\frac{2}{3}|H_{0}\rangle|H_{0}\rangle+\frac{2}{3}|V_{1}\rangle|V_{1}\rangle
+\frac{1}{3}|H_{2}\rangle|H_{2}\rangle)\otimes|H\rangle_{t} \\
=&(\frac{2}{3}\alpha|H_{0}\rangle|H_{0}\rangle|H_{0}\rangle
+\frac{2}{3}\alpha|H_{0}\rangle|V_{1}\rangle|V_{1}\rangle
+\frac{1}{3}\alpha|H_{0}\rangle|H_{2}\rangle|H_{2}\rangle \\
&+\frac{2}{3}\beta|V_{1}\rangle|H_{0}\rangle|H_{0}\rangle
+\frac{2}{3}\beta|V_{1}\rangle|V_{1}\rangle|V_{1}\rangle
+\frac{1}{3}\beta|V_{1}\rangle|H_{2}\rangle|H_{2}\rangle \\
&+\frac{2}{3}\gamma|H_{2}\rangle|H_{0}\rangle|H_{0}\rangle
+\frac{2}{3}\gamma|H_{2}\rangle|V_{1}\rangle|V_{1}\rangle
+\frac{1}{3}\gamma|H_{2}\rangle|H_{2}\rangle|H_{2}\rangle)\otimes|H\rangle_{t}. \\
\end{split}
\end{equation}

When photons 1 and 2 pass through PBS1, we neglect the instance when two photons go to the same side and hence obtain the quantum state,
\begin{equation}
\begin{split}
(\frac{2}{3}\alpha|H_{0}\rangle_{a}|H_{0}\rangle_{b}|H_{0}\rangle
+\frac{1}{3}\alpha|H_{0}\rangle_{a}|H_{2}\rangle_{b}|H_{2}\rangle
+\frac{2}{3}\beta|V_{1}\rangle_{a}|V_{1}\rangle_{b}|V_{1}\rangle +\frac{2}{3}\gamma|H_{2}\rangle_{a}|H_{0}\rangle_{b}|H_{0}\rangle
+\frac{1}{3}\gamma|H_{2}\rangle_{a}|H_{2}\rangle_{b}|H_{2}\rangle)\otimes|H\rangle_{t}.
\end{split}
\end{equation}

The label "a" and "b" stand for beam splitter output ports.

Then, after BD1 and BD3, the quantum state becomes
\begin{equation}
\begin{split}
(\frac{2}{3}\alpha|V_{0}\rangle_{a}|V_{0}\rangle_{b}|H_{0}\rangle
+\frac{1}{3}\alpha|V_{0}\rangle_{a}|H_{1}\rangle_{b}|H_{2}\rangle
+\frac{2}{3}\beta|H_{0}\rangle_{a}|H_{0}\rangle_{b}|V_{0}\rangle+\frac{2}{3}\gamma|H_{1}\rangle_{a}|V_{0}\rangle_{b}|H_{0}\rangle
+\frac{1}{3}\gamma|H_{1}\rangle_{a}|H_{1}\rangle_{b}|H_{1}\rangle)\otimes|H\rangle_{t}.
\end{split}
\end{equation}
After the HWPs, the quantum state becomes
\begin{equation}
\begin{split}
(\frac{2}{3}\alpha\frac{1}{\sqrt{2}}|H_{0}-V_{0}\rangle_{a}\frac{1}{\sqrt{2}}
|H_{0}-V_{0}\rangle_{b}|H_{0}\rangle
+\frac{1}{3}\alpha\frac{1}{\sqrt{2}}|H_{0}-V_{0}\rangle_{a}|H_{1}\rangle_{b}|H_{2}\rangle
+\frac{2}{3}\beta\frac{1}{\sqrt{2}}|H_{0}
+V_{0}\rangle_{a}\frac{1}{\sqrt{2}}|H_{0}+V_{0}\rangle_{b}|V_{1}\rangle\\
+\frac{2}{3}\gamma|H_{1}\rangle_{a}\frac{1}{\sqrt{2}}|H_{0}-V_{0}\rangle_{b}|H_{0}\rangle
+\frac{1}{3}\gamma|H_{1}\rangle_{a}|H_{1}\rangle_{b}|H_{2}\rangle)\otimes|H\rangle_{t}.
\end{split}
\end{equation}
After BD2 and BD4, the quantum state becomes
\begin{equation}
\begin{split}
(\frac{1}{3}\alpha|H_{0}\rangle_{a}|H_{0}\rangle_{b}|H_{0}\rangle
+\frac{\sqrt{2}}{6}\alpha|H_{0}\rangle_{a}|V_{0}\rangle_{b}|H_{2}\rangle +\frac{1}{3}\beta|H_{0}\rangle_{a}|H_{0}\rangle_{b}|V_{1}\rangle
+\frac{\sqrt{2}}{3}\gamma|V_{0}\rangle_{a}|H_{0}\rangle_{b}|H_{0}\rangle
+\frac{1}{3}\gamma|V_{0}\rangle_{a}|V_{0}\rangle_{b}|H_{2}\rangle)\otimes|H\rangle_{t}.
\end{split}
\end{equation}

To remove the two noise terms, we introduce auxiliary entanglement $|\varepsilon\rangle_{45}=(|H\rangle_{c}|H\rangle_{d}+|V\rangle_{c}|V\rangle_{d})/\sqrt{2}$ on photon 4--5.

After PBS2 and PBS3, we only retain the six-fold coincidence quantum states
\begin{equation}
\begin{split}
&\frac{\sqrt{2}}{6}(\alpha|H\rangle_{a}|H\rangle_{b}|H\rangle_{c}|H\rangle_{d}|H_{0}\rangle
+\beta|H\rangle_{a}|H\rangle_{b}|H\rangle_{c}|H\rangle_{d}|V_{1}\rangle
+\gamma|V\rangle_{a}|V\rangle_{b}|V\rangle_{c}|V\rangle_{d}|H_{2}\rangle)\otimes|H\rangle_{t}.
\end{split}
\end{equation}
After setting HWP1--4 to $22.5^\circ$, the quantum state changes to
\begin{equation}
\begin{split}
(&\frac{1}{4}(|HHHH\rangle+|HHVV\rangle+|HVHV\rangle+|HVVH\rangle \\
&+|VHHV\rangle+|VHVH\rangle+|VVHH\rangle+|VVVV\rangle)\frac{\sqrt{2}}{6}(\alpha|H_{0}\rangle+\beta|V_{1}\rangle+\gamma|H_{2}\rangle) \\
&+\frac{1}{4}(|HHHV\rangle+|HHVH\rangle+|HVHH\rangle+|HVVV\rangle \\
&+|VHHH\rangle+|VHVV\rangle+|VVHV\rangle+|VVVH\rangle)\frac{\sqrt{2}}{6}(\alpha|H_{0}\rangle+\beta|V_{1}\rangle
-\gamma|H_{2}\rangle))\otimes|H\rangle_{t}.
\end{split}
\end{equation}

Finally, we completed the teleportation according to the six-fold coincidence between parts 1, 2, 3, 4 (see Fig. \ref{fig10}) and trigger photon t, photon 3. The probability of a successful teleportation is 1/18.\\

\begin{figure}[tbph]
\begin{center}
\includegraphics [width= 0.5\columnwidth]{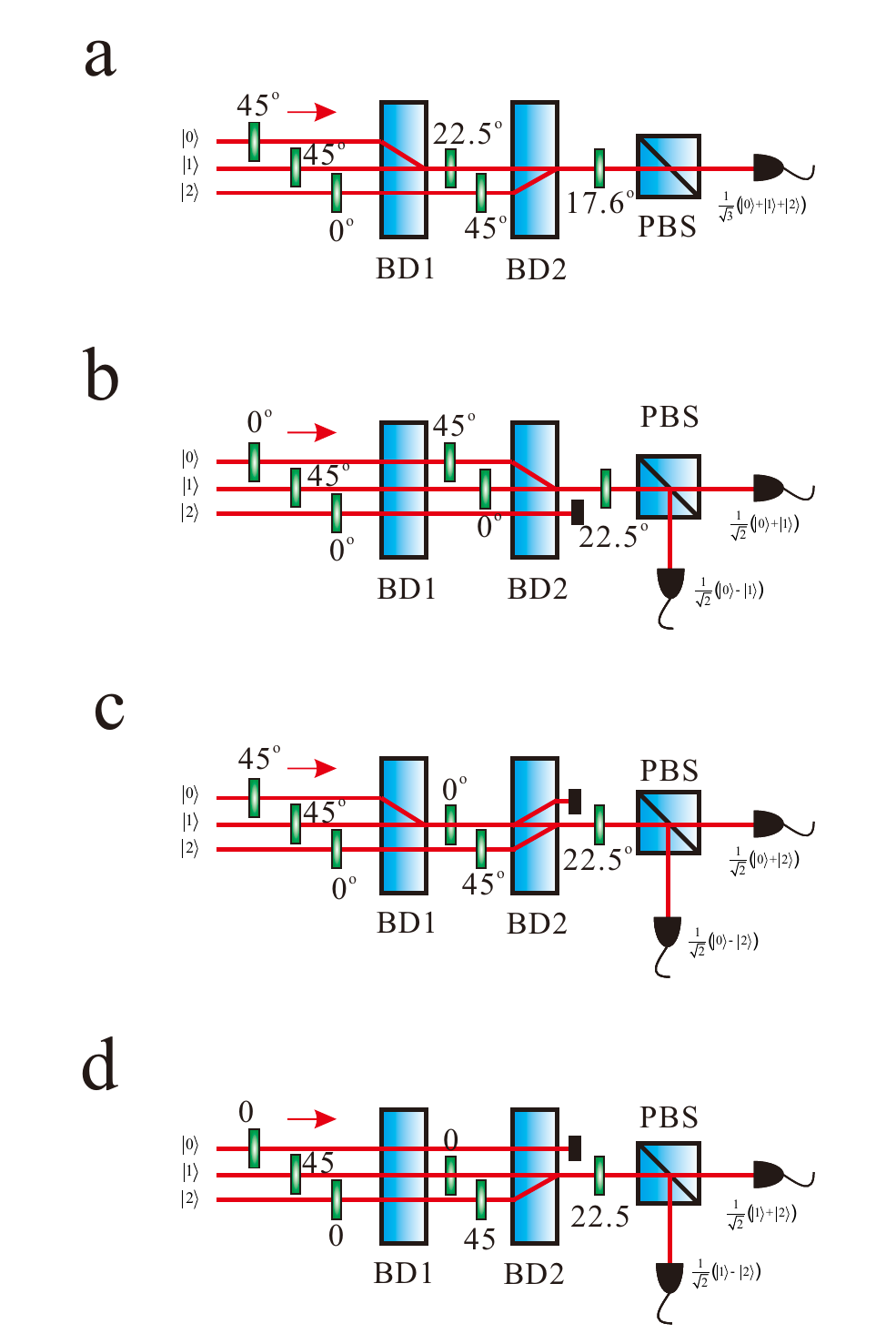}
\end{center}
\caption{\textbf{Typical single-observable measurement devices.} Experimental setups for measuring: \textbf{a}, $(|0\rangle+|1\rangle+|2\rangle)/\sqrt{3}$, \textbf{b}, $(|0\rangle\pm|1\rangle)/\sqrt{2}$, \textbf{c}, $(|0\rangle\pm|2\rangle)/\sqrt{2}$, and \textbf{d}, $(|1\rangle\pm|2\rangle)/\sqrt{2}$. Here, $|0\rangle$ and $|2\rangle$ are horizontally polarized, and $|1\rangle$ is vertically polarized.}
\label{fig11}
\vspace{-0.5cm}
\end{figure}

\noindent\textbf{Single photon measurement on photon 3.} For photon 3, we use different single-observable measurements to determine the fidelity (see Fig. \ref{fig11}).\\

\begin{figure}[tbph]
\begin{center}
\includegraphics [width= 0.5\columnwidth]{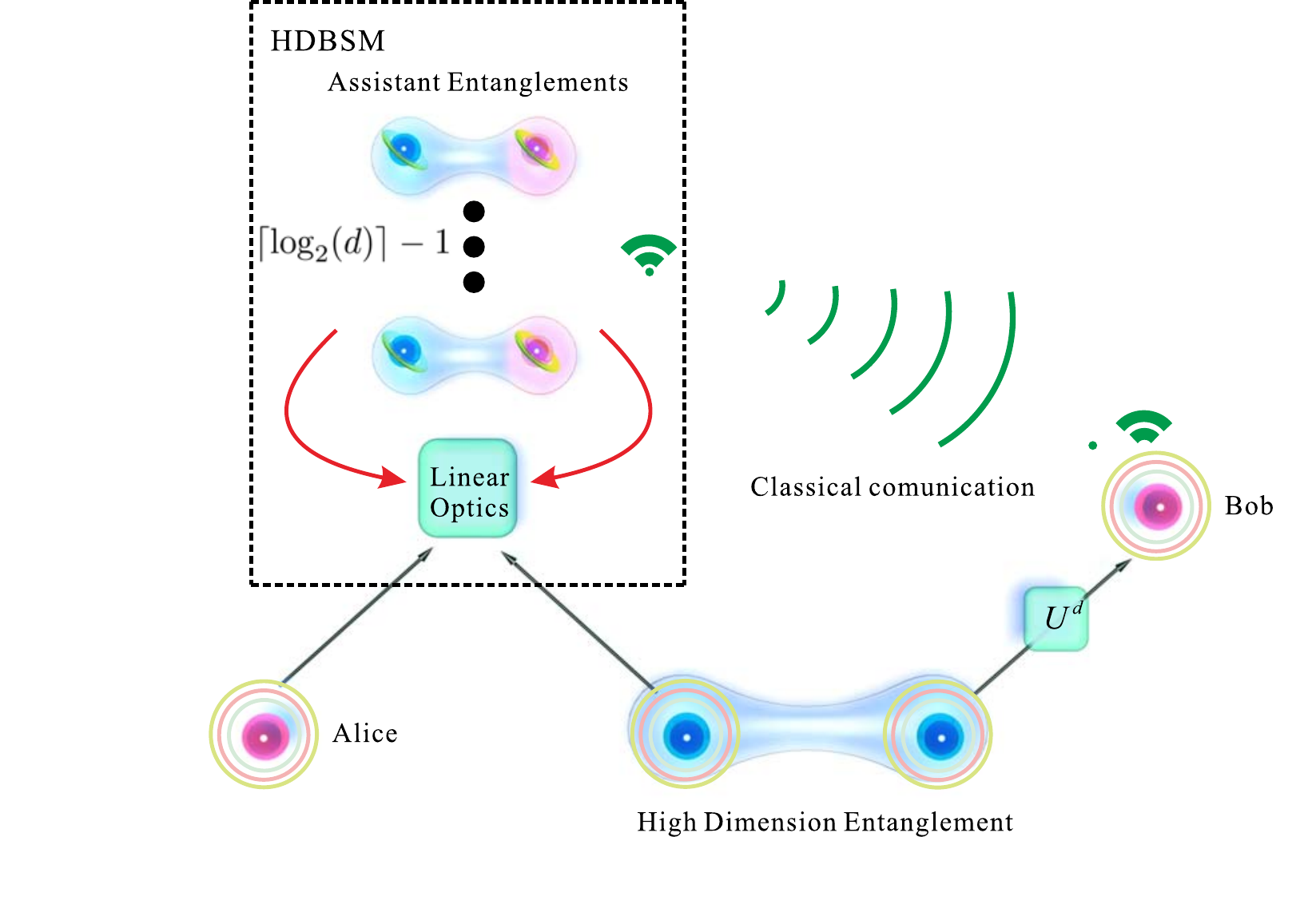}
\end{center}
\caption{\textbf{Schematic of general d-dimensional quantum state teleportation using linear optics.} Alice and Bob initially share a maximally entangled quantum state in d-dimension. The d-dimensional BSM is achieved by utilising additional $\lceil\log_{2}(d)\rceil-1$ auxiliary entangled photon  pairs. Bob completes the whole operation by performing the appropriate unitary operation based on the results of Alice's HDBSM.}
\label{fig12}
\vspace{-0.5cm}
\end{figure}

\noindent\textbf{Generalised protocol and experimental scheme for teleporting a d-dimensional quantum state.} In general, Alice wants to teleport a d-dimensional state (qudit) $|\varphi\rangle_{\mathrm{1}}=\sum_{i=0}^{d-1} \alpha_{i}|i\rangle_{\mathrm{1}}$ to Bob. Alice and Bob share a pair of photons in the state $|\psi\rangle_{\mathrm{23}}=\frac{1}{\sqrt{d}} \sum_{i=0}^{d-1}|i\rangle_{\mathrm{2}}|i\rangle_{\mathrm{3}}$, which is a maximally entangled state in d-dimension as shown in Fig. \ref{fig12}. Alice holds one photon of the entangled pair (photon 2), and Bob holds the other one (photon 3). Alice performs a d-dimension Bell state measurement on photons 1 and 2. In the basis:
\begin{equation}
\left|\psi_{n m}\right\rangle_{\mathrm{12}}=\frac{1}{\sqrt{d}} \sum_{j=0}^{d-1} e^{j 2 \pi j n / d}|j\rangle \otimes|(j+m) \bmod d\rangle.
\end{equation}

According to the results of Alice's HDBSM, Bob performs the corresponding unitary operation
\begin{equation}
U_{n m}=\sum_{k=0}^{d-1} e^{i 2 \pi k n / d}|k\rangle \otimes\langle(k+m) \bmod d|.
\end{equation}

Teleportation of d-dimensional states can be implemented by the above protocol.

For teleportation of d-dimension quantum states using linear optics, the challenge is to realize HDBSM. To realize a d-dimensional BSM, we need additional entangled photon pairs. For d-dimensional systems, the number of auxiliary entangled pairs we needed is $\lceil\log_{2}(d)\rceil-1$. We use the previous scheme, and the higher-dimensional solution is just a simple extension. In the first step, we control the coding of each path appropriately, convert even coding ($|0\rangle, |2\rangle, |4\rangle ...$) into H polarization and odd coding ($|1\rangle, |3\rangle, |5\rangle ...$) into V polarization. In the first step, we use polarizers to realize the post-selection of some classical items. Only the paths with the same polarization will have two-fold coincidence, and other classical items ($|i\rangle|i+1\rangle$) will be ignored. However, such a post-selection is not enough for us to select a deterministic high-dimensional Bell state. Our goal is to preserve the classical items ($|i\rangle|i\rangle$) and ignore all other classical items. However, for three-dimensional and four-dimensional Bell states, classical items such as $|i\rangle|i+2\rangle$ can not be excluded by post-selection of polarizers. So we introduce an entangled photon pairs to form a four-fold coincidence, excluding the classical term $|i\rangle|i+2\rangle$. Finally, we can identify a high-dimensional Bell state deterministically according to the phase information. By analogy, if we need to exclude the classical $|i\rangle|i+4\rangle$ items, we need to add another entangled photon pairs to form a six-fold coincidence. Then d = 5, 6, 7, 8-dimensional BSM can be completed. If we need to complete the teleportation of d-dimensional quantum states, we need $\lceil\log_{2}(d)\rceil-1$ auxiliary entangled pairs. \\

\noindent\textbf{The result of qutrit Tomography.} We reconstruct the density matrix of $|\psi_{1}\rangle-|\psi_{10}\rangle$ by using the standard qutrit state tomography. The reconstructed density matrix($\rho_{1}-\rho_{10}$) results are as follows:
\begin{equation}
\rho_{1}=\left(
\begin{array}{ccc}
0.745 + 0.000i & 0.110 - 0.109i & 0.075 - 0.053i\\
0.110 + 0.103i & 0.142 + 0.000i & 0.043 - 0.050i\\
0.075 + 0.053i & 0.043 + 0.050i & 0.114 + 0.000i\\
\end{array}
\right),
\end{equation}
\begin{equation}
\rho_{2}=\left(
\begin{array}{ccc}
   0.151 + 0.000i & -0.033 + 0.051i &  0.051 - 0.003i \\
  -0.033 - 0.051i  & 0.715 + 0.000i & -0.078 - 0.072i \\
   0.051 + 0.003i & -0.078 + 0.072i &  0.133 + 0.000i \\
\end{array}
\right),
\end{equation}
\begin{equation}
\rho_{3}=\left(
\begin{array}{ccc}
0.154 + 0.000i &  0.026 + 0.014i & -0.008 - 0.092i \\
0.026 - 0.014i &  0.1369 + 0.000i &  0.059 + 0.006i  \\
-0.008 + 0.092i &  0.059 - 0.006i &  0.708 + 0.000i  \\
\end{array}
\right),
\end{equation}
\begin{equation}
\rho_{4}=\left(
\begin{array}{ccc}
0.425 + 0.000i &  0.316 + 0.156i &  0.018 + 0.029i \\
0.316 - 0.156i  & 0.430 + 0.000i & -0.079 + 0.069i  \\
0.018 - 0.029i & -0.079 - 0.069i &  0.143 + 0.000i  \\
\end{array}
\right),
\end{equation}
\begin{equation}
\rho_{5}=\left(
\begin{array}{ccc}
   0.468 + 0.000i & -0.273 - 0.252i & -0.080 + 0.059i \\
  -0.273 + 0.252i &  0.411 + 0.000i & -0.010 + 0.000i \\
  -0.080 - 0.059i & -0.010 - 0.000i  & 0.119 + 0.000i \\
\end{array}
\right),
\end{equation}
\begin{equation}
\rho_{6}=\left(
\begin{array}{ccc}
   0.464 + 0.000i & -0.041 + 0.065i &  0.232 + 0.148i \\
  -0.041 - 0.065i &  0.142 + 0.000i & -0.035 - 0.017i \\
   0.232 - 0.148i & -0.035 + 0.017i &  0.392 + 0.000i \\
\end{array}
\right),
\end{equation}
\begin{equation}
\rho_{7}=\left(
\begin{array}{ccc}
 0.418 + 0.000i &  0.009 + 0.049i & -0.221 - 0.192i \\
   0.009 - 0.049i &  0.133 + 0.000i & -0.064 - 0.014i \\
  -0.221 + 0.192i & -0.064 + 0.014i &  0.448 + 0.000i  \\
\end{array}
\right),
\end{equation}
\begin{equation}
\rho_{8}=\left(
\begin{array}{ccc}
 0.158 + 0.000i &  0.000 + 0.051i & -0.053 + 0.043i \\
   0.000 - 0.051i &  0.362 + 0.000i &  0.247 + 0.140i \\
  -0.053 - 0.043i &  0.247 - 0.140i &  0.479 + 0.000i  \\
\end{array}
\right),
\end{equation}
\begin{equation}
\rho_{9}=\left(
\begin{array}{ccc}
 0.159 + 0.000i  & 0.005 + 0.065i & -0.074 - 0.008i \\
   0.005 - 0.065i &  0.390 + 0.000i & -0.079 - 0.244i \\
  -0.074 + 0.008i & -0.079 + 0.244i &  0.450 + 0.000i \\
\end{array}
\right),
\end{equation}
\begin{equation}
\rho_{10}=\left(
\begin{array}{ccc}
0.321 + 0.000i &  0.111 + 0.047i &  0.133 - 0.054i  \\
   0.111 - 0.047i  & 0.386 + 0.000i &  0.225 - 0.002i  \\
   0.133 + 0.054i  & 0.225 + 0.002i  & 0.292 + 0.000i  \\
\end{array}
\right).
\end{equation}

The fidelity of $\rho_{1}-\rho_{10}$ is in sequence:  $0.745\pm0.028$, $0.715\pm0.038$, $0.708\pm0.035$, $0.724\pm0.055$, $ 0.693\pm 0.065$, $0.661\pm0.071$, $0.626\pm0.062$, $0.668\pm0.006$, $0.643\pm0.037$, $0.665\pm0.055$, and $0.647\pm0.075$. The average fidelity of these 10 states is $F=0.685\pm0.022$. \\

\noindent\textbf{Qutrit process tomography.} By reconstructing the process matrix, we can fully describe the whole qutrit teleportation. The nine input states ($\varphi_{1}-\varphi_{9}$) are transferred to the corresponding output states ($\rho_{1}-\rho_{9}$). We can completely describe the effect of teleportation on the input states ($\rho_{ideal}$) by determining the process matrix $\chi$, defined by $\rho=\sum_{l,k=0}^{8}\chi_{l k}\sigma_{l} \rho_{ideal}\sigma_{k}$, where the $\sigma_{i}$ are the Pauli matrices:

\begin{equation}
\begin{split}
&\sigma_{1}=\left(
\begin{array}{ccc}
0 & 1 & 0\\
1 & 0 & 0\\
0 & 0 & 0\\
\end{array}
\right),\
\sigma_{2}=\left(
\begin{array}{ccc}
0 & -i & 0\\
i & 0 & 0\\
0 & 0 & 0\\
\end{array}
\right), \
\sigma_{3}=\left(
\begin{array}{ccc}
1 & 0 & 0\\
0 & -1 & 0\\
0 & 0 & 0\\
\end{array}
\right), \\
&\sigma_{4}=\left(
\begin{array}{ccc}
0 & 0 & 1\\
0 & 0 & 0\\
1 & 0 & 0\\
\end{array}
\right),\
\sigma_{5}=\left(
\begin{array}{ccc}
0 & 0 & -i\\
1 & 0 & 0\\
i & 0 & 0\\
\end{array}
\right), \
\sigma_{6}=\left(
\begin{array}{ccc}
0 & 0 & 0\\
0 & 0 & 1\\
0 & 1 & 0\\
\end{array}
\right), \\
&\sigma_{7}=\left(
\begin{array}{ccc}
0 & 0 & 0\\
0 & 0 & -i\\
0 & i & 0\\
\end{array}
\right),\
\sigma_{8}=\frac{1}{\sqrt{3}}\left(
\begin{array}{ccc}
1 & 0 & 0\\
0 & 1 & 0\\
0 & 0 & -2\\
\end{array}
\right). \\
\end{split}
\end{equation}

According to the results of $\rho_{1}-\rho_{9}$ state tomography, we can reconstruct the process matrix by maximum likelihood estimation.
\begin{tiny}
\begin{equation}
\begin{split}
&\chi=\left(
\begin{array}{ccccccccc}
   0.596 + 0.000i & -0.018 - 0.003i & -0.024 + 0.018i &  0.002 + 0.064i & -0.009 + 0.012i & -0.012 - 0.006i & -0.029 - 0.013i & 0.012 + 0.009i &  0.020 - 0.001i \\
  -0.018 + 0.003i &  0.059 + 0.000i & -0.045 - 0.007i &  0.000 - 0.016i &  0.010 - 0.006i & -0.008 + 0.011i &  0.006 + 0.003i & 0.000 - 0.011i & -0.001 + 0.022i \\
  -0.024 - 0.018i & -0.045 + 0.007i &  0.046 - 0.000i &  0.008 + 0.005i & -0.010 + 0.002i &  0.004 - 0.010i &  0.005 + 0.008i & -0.005 + 0.002i &  0.003 - 0.021i \\
   0.002 - 0.064i &  0.000 + 0.016i &  0.008 - 0.005i &  0.030 + 0.000i & -0.001 - 0.013i & -0.005 + 0.012i & -0.006 - 0.002i &   0.008 + 0.008i &  0.011 + 0.004i \\
  -0.009 - 0.012i &   0.010 + 0.006i & -0.010 - 0.002i &  -0.001 + 0.013i &   0.054 - 0.000i &  -0.041 - 0.006i & -0.000 + 0.008i &   -0.011 - 0.003i  & 0.022 - 0.014i \\
  -0.012 + 0.006i & -0.008 - 0.011i &  0.004 + 0.010i &  -0.005 - 0.012i & -0.041 + 0.006i  & 0.041 + 0.000i & -0.014 - 0.004i  &   0.018 - 0.001i & -0.013 + 0.015i  \\
  -0.029 + 0.013i &  0.006 - 0.003i &   0.005 - 0.008i &  -0.006 + 0.002i & -0.000 - 0.008i &  -0.014 + 0.004i &  0.052 + 0.000i &    -0.037 + 0.001i &  -0.029 - 0.002i  \\
   0.012 - 0.009i  & 0.000 + 0.011i &  -0.005 - 0.002i &  0.008 - 0.008i &  -0.011 + 0.003i &  0.018 + 0.001i &  -0.037 - 0.001i &      0.032 - 0.000i &  0.009 + 0.006i \\
   0.020 + 0.001i & -0.001 - 0.022i &  0.003 + 0.021i &  0.011 - 0.004i &  0.022 + 0.014i & -0.013 - 0.015i & -0.029 + 0.002i  &    0.009 - 0.006i  & 0.087 + 0.000i \\
\end{array}
\right).\\
\end{split}
\end{equation}
\end{tiny}
The ideal process matrix of quantum teleportation $\chi_{ideal}$ has only one non-zero component, $(\chi_{ideal})_{00}=1$, this represents the perfect teleportation process. The process fidelity of our experiment was $f_{\text {process }}=\operatorname{Tr}\left(\chi_{{ideal}} \chi\right)=0.596\pm0.037$.  For three-dimensional nonclassical teleportation, the bound of average fidelity is $f_{\text {stateave}}=0.5$, so the nonclassical bound of process matrix is $f_{\text {process }}=1/3$ according to the relation between the average state fidelity $f_{\text {stateave}}$ and the process fidelity \cite{Gilchrist2005}:
\begin{equation}
f_{\text {stateave}}= \frac{f_{\text {process }}d+1}{d+1},
\end{equation}
where d is the dimension of the quantum system. In our experiment, the process fidelity is $f_{\text {process }}=\operatorname{Tr}\left(\chi_{{ideal}} \chi\right)=0.596\pm0.037$, which exceeds the nonclassical bound by 7 standard deviations, and proves that our teleportation is nonclassical.\\

\noindent\textbf{The fidelity of MUB basis state.} The MUB basis are:
\begin{equation}
\begin{split}
&|\psi_{1}\rangle=|0\rangle, \\
&|\psi_{2}\rangle=|1\rangle, \\
&|\psi_{3}\rangle=|2\rangle, \\
&|\psi_{4}\rangle=\frac{1}{\sqrt{3}}(|0\rangle+|1\rangle+|2\rangle), \\
&|\psi_{5}\rangle=\frac{1}{\sqrt{3}}(|0\rangle+\omega|1\rangle+\omega^2|2\rangle), \\
&|\psi_{6}\rangle=\frac{1}{\sqrt{3}}(|0\rangle+\omega^2|1\rangle+\omega|2\rangle),  \\
&|\psi_{7}\rangle=\frac{1}{\sqrt{3}}(\omega|0\rangle+|1\rangle+|2\rangle),  \\
&|\psi_{8}\rangle=\frac{1}{\sqrt{3}}(|0\rangle+\omega|1\rangle+|2\rangle),   \\
&|\psi_{9}\rangle=\frac{1}{\sqrt{3}}(|0\rangle+|1\rangle+\omega|2\rangle),   \\
&|\psi_{10}\rangle=\frac{1}{\sqrt{3}}(\omega^2|0\rangle+|1\rangle+|2\rangle),   \\
&|\psi_{11}\rangle=\frac{1}{\sqrt{3}}(|0\rangle+\omega^2|1\rangle+|2\rangle),   \\
&|\psi_{12}\rangle=\frac{1}{\sqrt{3}}(|0\rangle+|1\rangle+\omega^2|2\rangle),
\end{split}
\end{equation}
where $\omega=exp(i2\pi/3)$. After the evolution of these ideal states ($|\psi_{1}\rangle-|\psi_{12}\rangle$) through the process matrix $\chi$, 12 output states are obtained ($\rho=\sum_{l,k=0}^{8}\chi_{l k}\sigma_{l}\rho_{\text {ideal}}\sigma_{k}$). Thus we can calculate the fidelity of each state, they are ($F_{1}-F_{12}$=$0.740\pm0.029$, $0.689\pm0.029$, $0.713\pm0.031$, $0.634\pm0.056$, $0.728\pm0.049$, $0.687\pm0.063$, $0.674\pm0.049$, $0.664\pm0.055$, $0.751\pm0.056$, $0.668\pm0.050$, $0.764\pm0.047$, $0.648\pm0.062$.) The average fidelity of these 12 states is $F=0.697\pm0.026$, which is significantly higher than the classical upper bound of 0.5.   \\

\noindent\textbf{The criteria for genuine qutrit state.} For high-dimensional teleportation, the first step is to confirm that teleportation is non-classical that is, it cannot be simulated classically. In the second step, we need to confirm the dimension of quantum states contained in the particles we teleported. In our qutrit case, we need to determine whether the state of teleportation is a coherent superposition of three levels or merely an incoherent mixture of qubit states. Intuitively, if the fidelity between the states we prepared and state $1/\sqrt{3}(|0\rangle+|1\rangle+|2\rangle)$ is more than 2/3, we can be sure that the prepared states are genuine qutrit states. However, this is a sufficient but not a necessary condition. For some states (like $\sqrt{1/8}|0\rangle+\sqrt{1/8}|1\rangle-\sqrt{3/4}|2\rangle$), the fidelity does not reach 2/3, they are nonetheless genuine three-dimensional coherent superposition states.

In this subsection, we describe how to certify the genuine high dimensionality of the teleported states. We first give the nonlinear criterion to determine the genuine three-dimensional state and prove that the fidelity of 2/3 is not a necessary and sufficient condition. Then we give the robustness of the genuine three-dimensional entanglement and prove that the teleportation we have completed is truly three-dimensional.

The density matrix of a qutrit can be described as
\begin{equation}
\rho_{qutrit}=\frac{1}{3}\left(\begin{array}{ccc}
1+\frac{\sqrt{3}}{2}(\langle\lambda_8\rangle+\sqrt{3}\langle\lambda_3\rangle) & \frac{3}{2}(\langle\lambda_1\rangle-i\langle\lambda_2\rangle) & \frac{3}{2}(\langle\lambda_4\rangle-i\langle\lambda_5\rangle)\vspace{1ex}\\
\frac{3}{2}(\langle\lambda_1\rangle+i\langle\lambda_2\rangle) & 1+\frac{\sqrt{3}}{2}(\langle\lambda_8\rangle-\sqrt{3}\langle\lambda_3\rangle) & \frac{3}{2}(\langle\lambda_6\rangle-i\langle\lambda_7\rangle)\vspace{1ex}\\
\frac{3}{2}(\langle\lambda_4\rangle-i\langle\lambda_5\rangle) & \frac{3}{2}(\langle\lambda_6\rangle+i\langle\lambda_7\rangle) & 1-\sqrt{3}\langle\lambda_8\rangle
\end{array}\right),
\end{equation}
where $\langle\lambda_i\rangle=Tr(\rho_{qutrit}\lambda_i)$ and
\begin{equation}
\begin{array}{cccc}
\lambda_1=\left[\begin{array}{ccc}
0 & 1 & 0 \\
1 & 0 & 0 \\
0 & 0 & 0
\end{array}\right], & \lambda_2=\left[\begin{array}{ccc}
0 & -i & 0 \\
i & 0 & 0 \\
0 & 0 & 0
\end{array}\right], & \lambda_3=\left[\begin{array}{ccc}
1 & 0 & 0 \\
0 & -1 & 0 \\
0 & 0 & 0
\end{array}\right], & \lambda_4=\left[\begin{array}{ccc}
0 & 0 & 1 \\
0 & 0 & 0 \\
1 & 0 & 0
\end{array}\right], \\
\\
\lambda_5=\left[\begin{array}{ccc}
0 & 0 & -i \\
0 & 0 & 0 \\
i & 0 & 0
\end{array}\right], & \lambda_6=\left[\begin{array}{ccc}
0 & 0 & 0 \\
0 & 0 & 1 \\
0 & 1 & 0
\end{array}\right], & \lambda_7=\left[\begin{array}{ccc}
0 & 0 & 0 \\
0 & 0 & -i \\
0 & i & 0
\end{array}\right], & \lambda_8=\frac{1}{\sqrt{3}}\left[\begin{array}{ccc}
1 & 0 & 0 \\
0 & 1 & 0 \\
0 & 0 & -2
\end{array}\right].
\end{array}
\end{equation}

Let us consider the convex combination of these subspaces: $\{|0\rangle,|1\rangle\}$, $\{|0\rangle,|2\rangle\}$ and $\{|1\rangle,|2\rangle\}$, $\rho_{qubit}=P_1\rho_{01}+P_2\rho_{02}+P_3\rho_{12}\quad \text{where}\quad 0\leq P_i\leq 1,\ \sum_i P_i=1, i=1,2,3. $

\begin{equation}
\rho_{01}=\frac{1}{2}\left(\begin{array}{ccc}
1 & 0 & 0\\
0 & 1 & 0\\
0 & 0 & 0
\end{array}\right)+\frac{x_{01}}{2}\left(\begin{array}{ccc}
0 & 1 & 0\\
1 & 0 & 0\\
0 & 0 & 0
\end{array}\right)+\frac{y_{01}}{2}\left(\begin{array}{ccc}
0 & -i & 0\\
i & 0 & 0\\
0 & 0 & 0
\end{array}\right)+\frac{z_{01}}{2}\left(\begin{array}{ccc}
1 & 0 & 0\\
0 & -1 & 0\\
0 & 0 & 0
\end{array}\right),\\
\end{equation}
\begin{equation}
\rho_{02}=\frac{1}{2}\left(\begin{array}{ccc}
1 & 0 & 0\\
0 & 0 & 0\\
0 & 0 & 1
\end{array}\right)+\frac{x_{02}}{2}\left(\begin{array}{ccc}
0 & 0 & 1\\
0 & 0 & 0\\
1 & 0 & 0
\end{array}\right)+\frac{y_{02}}{2}\left(\begin{array}{ccc}
0 & 0 & -i\\
0 & 0 & 0\\
i & 0 & 0
\end{array}\right)+\frac{z_{02}}{2}\left(\begin{array}{ccc}
1 & 0 & 0\\
0 & 0 & 0\\
0 & 0 & -1
\end{array}\right),\\
\end{equation}
\begin{equation}
\rho_{12}=\frac{1}{2}\left(\begin{array}{ccc}
0 & 0 & 0\\
0 & 1 & 0\\
0 & 0 & 1
\end{array}\right)+\frac{x_{12}}{2}\left(\begin{array}{ccc}
0 & 0 & 0\\
0 & 0 & 1\\
0 & 1 & 0
\end{array}\right)+\frac{y_{12}}{2}\left(\begin{array}{ccc}
0 & 0 & 0\\
0 & 0 & -i\\
0 & i & 0
\end{array}\right)+\frac{z_{12}}{2}\left(\begin{array}{ccc}
0 & 0 & 0\\
0 & 1 & 0\\
0 & 0 & -1
\end{array}\right),\\
\end{equation}
where $x_{i}^{2}+y_{i}^{2}+z_{i}^{2}\leq 1\quad \big( i\in \{01,02,12\}\big)$.\\

Note that $\rho_{01}$, $\rho_{02}$ and  $\rho_{12}$ only has a coherent superposition of two levels, any experiment output of $\rho_{qubit}$ can be simulated by qubit systems. Thus $\rho_{qubit}$ is not a genuine qutrit.

Now let's find the criteria for genuine qutrit states. A general qutrit $\rho_{qutrit}$ which can be simulated by $\rho_{qubit}$ means:
\begin{equation}
\rho_{qutrit}=\rho_{qubit}.
\label{equ31}
\end{equation}
To solve equation~(\ref{equ31}) is equivalent to solve the following equations
\begin{equation}
\begin{array}{ccc}
\left\{
\begin{aligned}
    P_{1}x_{01}=\langle\lambda_1\rangle\\
    P_{1}y_{01}=\langle\lambda_2\rangle\\
\end{aligned},
\right .& \left\{
\begin{aligned}
    P_{2}x_{02}=\langle\lambda_4\rangle\\
    P_{2}y_{02}=\langle\lambda_5\rangle\\
\end{aligned},
\right .& \left\{
\begin{aligned}
    P_{3}x_{12}=\langle\lambda_6\rangle\\
    P_{3}y_{12}=\langle\lambda_7\rangle,\\
\end{aligned}
\right .
\end{array}
\label{equ32}
\end{equation}
\begin{eqnarray}
\left\{
\begin{aligned}
P_1+P_2 z_{02}+P_3 z_{12}&=\frac{1}{3}+\frac{2\langle\lambda_8\rangle}{\sqrt{3}}\\
P_1z_{01}+P_2+P_3 (-z_{12})&=\frac{1}{3}+\langle\lambda_3\rangle-\frac{\langle\lambda_8\rangle}{\sqrt{3}}\\
P_1z_{01}+P_2 z_{02}+P_3 (-1)&=-\frac{1}{3}+\langle\lambda_3\rangle+\frac{\langle\lambda_8\rangle}{\sqrt{3}}\\
\end{aligned}
\right .
\end{eqnarray}

Note that $x_i,\ y_i,\ z_i \in[-1,1]\quad (i\in\{01,02,12\})\ \text{and}\ \sum\limits_{i=1}^3 P_i=1$, equation~(\ref{equ32}) gives a group of linear criteria:\\
\begin{equation}
|\langle\lambda_1\rangle+\langle\lambda_4\rangle+\langle\lambda_6\rangle|\leq1,
\label{equ34}
\end{equation}
\begin{equation}
|\langle\lambda_1\rangle+\langle\lambda_4\rangle+\langle\lambda_7\rangle|\leq1,
\label{equ35}
\end{equation}
\begin{equation}
|\langle\lambda_1\rangle+\langle\lambda_5\rangle+\langle\lambda_6\rangle|\leq1,
\label{equ36}
\end{equation}
\begin{equation}
|\langle\lambda_1\rangle+\langle\lambda_5\rangle+\langle\lambda_7\rangle|\leq1,
\label{equ37}
\end{equation}
\begin{equation}
|\langle\lambda_2\rangle+\langle\lambda_4\rangle+\langle\lambda_6\rangle|\leq1,
\label{equ38}
\end{equation}
\begin{equation}
|\langle\lambda_2\rangle+\langle\lambda_4\rangle+\langle\lambda_7\rangle|\leq1,
\label{equ39}
\end{equation}
\begin{equation}
|\langle\lambda_2\rangle+\langle\lambda_5\rangle+\langle\lambda_6\rangle|\leq1,
\label{equ40}
\end{equation}
\begin{equation}
|\langle\lambda_2\rangle+\langle\lambda_5\rangle+\langle\lambda_7\rangle|\leq1.
\label{equ41}
\end{equation}

In Ref. \cite{Luo2019}, the authors present a fidelity criterion: any $\rho_{qubit}$ must satisfy $Tr[\rho_{qubit}|\psi\rangle\langle\psi |]\leq\frac{2}{3}$, where $|\psi\rangle=\frac{1}{\sqrt{3}}(|0\rangle+|1\rangle+|2\rangle)$. Note that their criterion is equivalent to our linear criterion~(\ref{equ34}). There are some genuine qutrit states can be detected by criterion~(\ref{equ35}) while can not be detected by the fidelity criteria here (nor by other fidelity criteria in their paper).

Now let's present a more powerful nonlinear criterion. Any $\rho_{qubit}$ must satisfy:
\begin{equation}
\sqrt{\langle\lambda_1\rangle^2+\langle\lambda_2\rangle^2}+\sqrt{\langle\lambda_4\rangle^2+\langle\lambda_5\rangle^2}+\sqrt{\langle\lambda_6\rangle^2+\langle\lambda_7\rangle^2}\leq1.
\label{equ42}
\end{equation}

The numerical result shows this criterion is more powerful than the fidelity criterion, which means any states testable by the fidelity criterion can be test by criterion~(\ref{equ42}), but not vice versa.

The proof of criterion~(\ref{equ42}) is similar to the proof of criteria~(\ref{equ34})-(\ref{equ41}). $(2a)^2+(2b)^2$ gives:
\begin{equation}
P_1^2(x_{01}^2+y_{01}^2)=\langle\lambda_1\rangle^2+\langle\lambda_2\rangle^2,
\end{equation}
thus
\begin{equation}
P_1\sqrt{(x_{01}^2+y_{01}^2)}=\sqrt{\langle\lambda_1\rangle^2+\langle\lambda_2\rangle^2}.\\
\end{equation}
Similarly, we have:
\begin{eqnarray}
\qquad &P_2\sqrt{(x_{02}^2+y_{02}^2)}=\sqrt{\langle\lambda_4\rangle^2+\langle\lambda_5\rangle^2},\\
&P_3\sqrt{(x_{12}^2+y_{12}^2)}=\sqrt{\langle\lambda_6\rangle^2+\langle\lambda_7\rangle^2}.
\end{eqnarray}
Notice that $\sqrt{x_i^2+y_i^2}\leq1\quad(i\in\{01,02,12\})$, we get criterion~(\ref{equ42}). By using this nonlinear criterion, we can judge that $\sqrt{1/8}|0\rangle+\sqrt{1/8}|1\rangle-\sqrt{3/4}|2\rangle$ is a genuine three-dimensional state, but the fidelity between this state and state $1/\sqrt{3}(|0\rangle+|1\rangle+|2\rangle)$ is less than 2/3.

The nonlinear criterion given above is not necessary and sufficient condition. We give the necessary and sufficient conditions of genuine three-dimensional state by semidefinite program (SDP). Testing if a state $\rho$ is a genuine qutrit can be cast into the following feasibility SDP:
\begin{align}
	\min_{\sigma_{ij}} & \quad 0 \nonumber \\
	 \text{subject to } & \quad \rho_{ij} \geq 0, \nonumber \\
		& \sum \operatorname{Tr}\left[\sigma_{i j}\right]=1 \\
			& \quad \rho = \sigma_{01}+\sigma_{12}+\sigma_{02}.
\end{align}
Note that $\sigma_{ij}$ is simply $p_{ij} \rho_{ij}$. If the above program is not feasible, then the state $\rho$ has to be a genuine qutrit state. Alternatively, one can consider a variational version of the feasibility program:
\begin{align}
	\mu:= \min_{\sigma_{ij}, \mu^*} & \quad \mu^*  \nonumber \\
			 \text{subject to} & \quad \sigma_{ij} \geq 0, \nonumber \\
		& \sum \operatorname{Tr}\left[\sigma_{i j}\right]=1 \\
             &\mu^* \frac{\mathbb{I}}{3}+(1-\mu^*) \rho=\sigma_{01}+\sigma_{12}+\sigma_{02}.
\end{align}
$\mu>0$ implies the first SDP is infeasible, hence $\rho$ is a genuine qutrit state. Whereas $\mu\leq0$ implies that $\rho$ is an incoherent mixture of two-level states. The value of $\mu$ can be interpreted as robustness of genuine dimensionality against white noise.

We choose a series of ideal state $(|0\rangle+e^{i\varphi_{1}}|1\rangle+e^{i\varphi_{2}}|2\rangle)/\sqrt{3}$ where $\varphi_{1}$, $\varphi_{2}$ are 20 phases at equal interval in $[0, \pi]$, and then get the states of teleportation through the process matrix evolution. We do the above SDP for these 400 states, 251 of which have robustness $\mu>0$. The average of these values is $\mu=0.111\pm0.034>0$, which can be successfully simulated outside the range of three standard deviations, it is proved that our teleportation is beyond qubit.\\

\noindent\textbf{Error propagation.} All the statistical error in our paper come from Monte Carlo method. In this section, we introduce the error propagation process of average fidelity and genuine three-dimensional robustness in detail.

\begin{figure}[tbph]
\begin{center}
\includegraphics [width= 0.5\columnwidth]{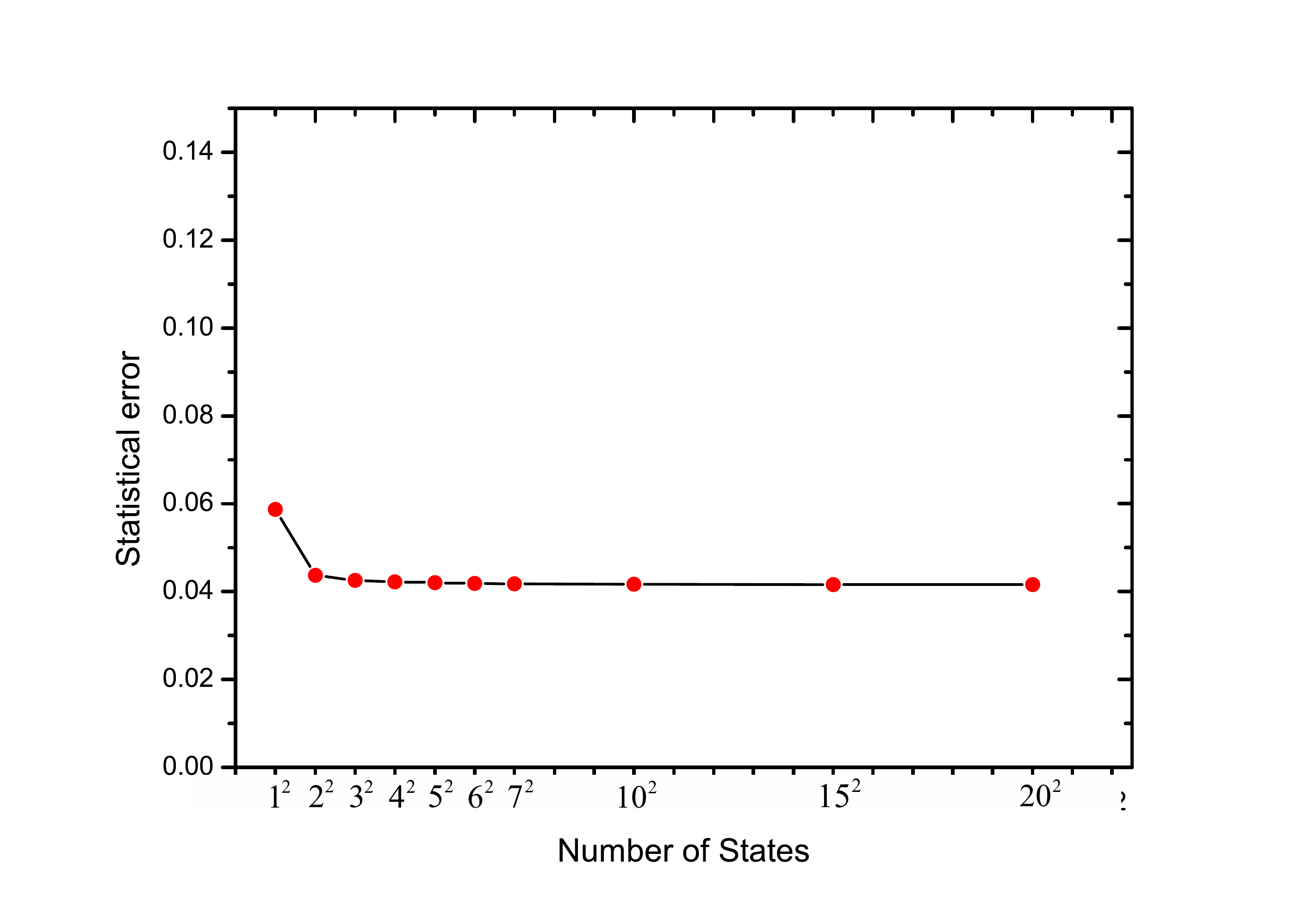}
\end{center}
\caption{\textbf{Convergence curve of statistical error of average fidelity.} A series of ideal state $1/\sqrt{3}(|0\rangle+e^{i\varphi_{1}}|1\rangle+e^{i\varphi_{2}}|2\rangle)$ are evenly selected, where $\varphi_{1},\varphi_{2}\in{[0,\pi]}$. The x-axis represents the number of sampling states. The y-axis represents the statistical error of average fidelity.}
\label{fig13}
\vspace{-0.5cm}
\end{figure}

Firstly, we generate 500 setting Poisson distribution simulation data based on raw data. Then we use these 500 sets of data to reconstruct 500 process matrices. These 500 process matrices obey Poisson statistical distribution. In order to obtain the statistical error of the average fidelity of 12 MUBs, we use 12 MUBs to input 500 process matrices, so we can get 500 average fidelity. We take standard deviation of 500 value as the statistical error of the 12 MUBs average fidelity. One may question that with the input state increases, due to the error propagation formula, the error of the average fidelity will be smaller and smaller.
\begin{equation}
S_{N}=\sqrt{\left(\frac{\partial f}{\partial x_{1}}\right)^{2} S_{x_{1}}^{2}+\left(\frac{\partial f}{\partial x_{2}}\right)^{2} S_{x_{2}}^{2}+\left(\frac{\partial f}{\partial x_{3}}\right)^{2} S_{x_{3}}^{2}},
\end{equation}
where $S$ stands for statistical error, $f$ stands for function, and $x$ stands for variable. However, the premise of this error propagation formula is that each variable $"x"$ is independent. In our case, the fidelity of the output state is affected by the process matrix. With the increase of the input state, the average fidelity will not be infinitely smaller, but will converge to a certain value. To verify this, We observed the statistical error of the average fidelity with the input state increased. As shown in Fig.~\ref{fig13}, we find that the statistical error converges rapidly to a certain value.

\begin{figure}[tbph]
\begin{center}
\includegraphics [width= 0.5\columnwidth]{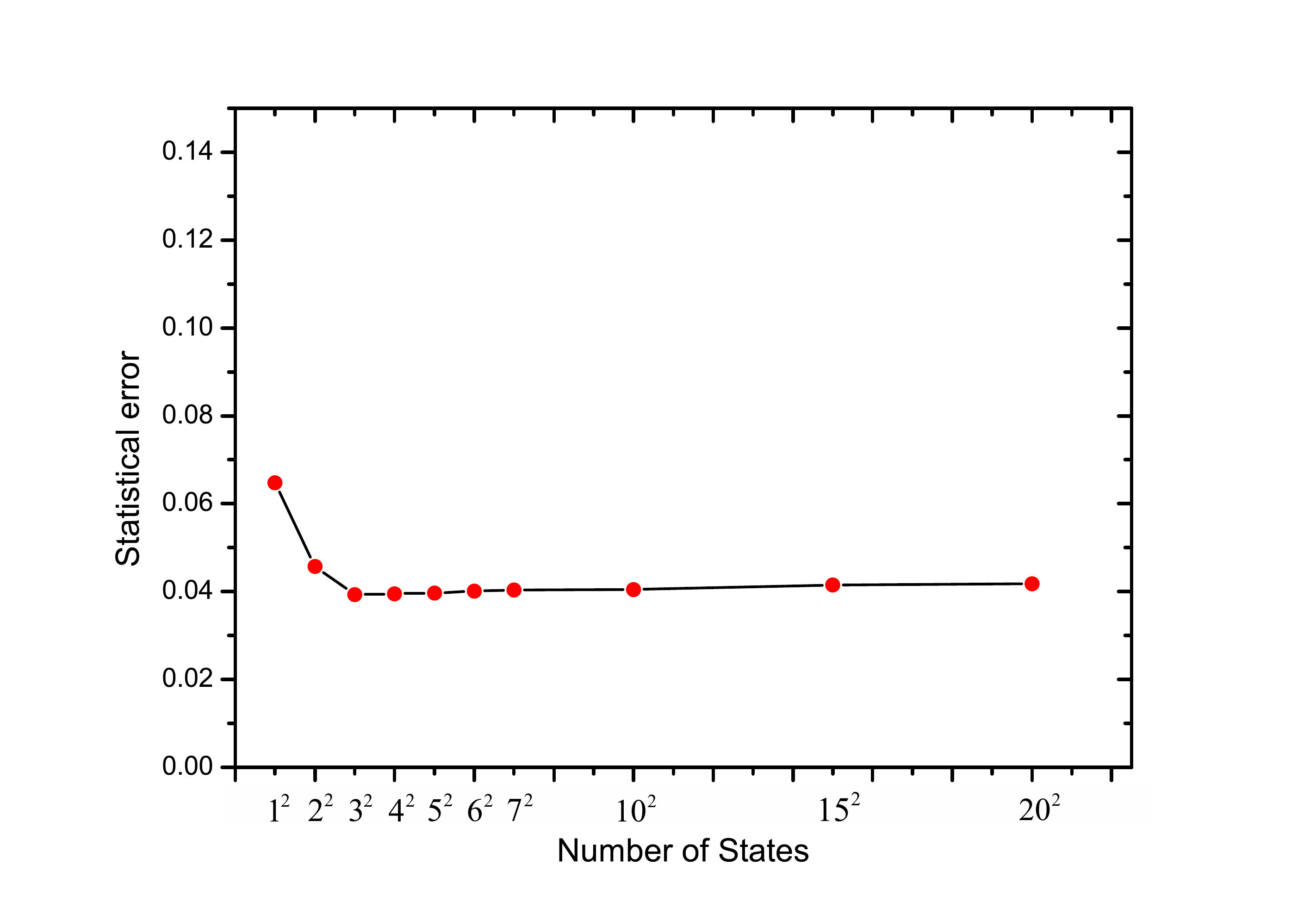}
\end{center}
\caption{\textbf{Convergence curve of statistical error of genuine three-dimensional robustness.} A series of ideal state $1/\sqrt{3}(|0\rangle+e^{i\varphi_{1}}|1\rangle+e^{i\varphi_{2}}|2\rangle)$ are selected, where $\varphi_{1},\varphi_{2}\in{[0,\pi]}$. The x-axis represents the number of sampling states. The y-axis represents the statistical error of genuine three-dimensional robustness.}
\label{fig14}
\vspace{-0.5cm}
\end{figure}

For the genuine three-dimensional robustness, the statistical error is estimated in the same way. Similarly, as shown in Fig.~\ref{fig14}, with the input state increases, the statistical error of robustness converges to a certain value.

One may question that the content of the process matrix reconstructed with 12 MUBs is different from that of the process matrix reconstructed with 9 bases we used. Some researches have shown that using MUBs can improve state estimation \cite{Adamson10}. However, the influence is subtle. It is still an open question in the case of process matrix and high dimension system. This subtle effect comes from the statistical fluctuation of the experiment. To study the influence of MUBs and non-MUBs reconstruction process density matrix on our experimental results, we can use the numerical simulation of Poisson distribution to do simulations of teleportation experiments.

Our numerical simulation uses 12 MUBs and 9 bases (used in our experiments) to reconstruct the process matrix respectively. Since the noise in our experiment is close to white noise, we assume that the process matrix of the three-dimensional channel in our experiment is:
\begin{equation}
\chi=0.55\chi_{ideal}+0.45\times I/9.
\end{equation}
We can use this process matrix $\chi$ to obtain the experiment measurement probability distribution without statistical fluctuation. Both MUB and non-MUB measurement probability distribution without data fluctuation can perfectly reconstruct the process matrix. Since the estimation accuracy of the two methods comes from the error caused by the statistical fluctuation of the data. We used Monte Carlo method to do 100 simulation experiments.

\begin{figure*}[tbph]
\begin{center}
\includegraphics [width= 0.5\columnwidth]{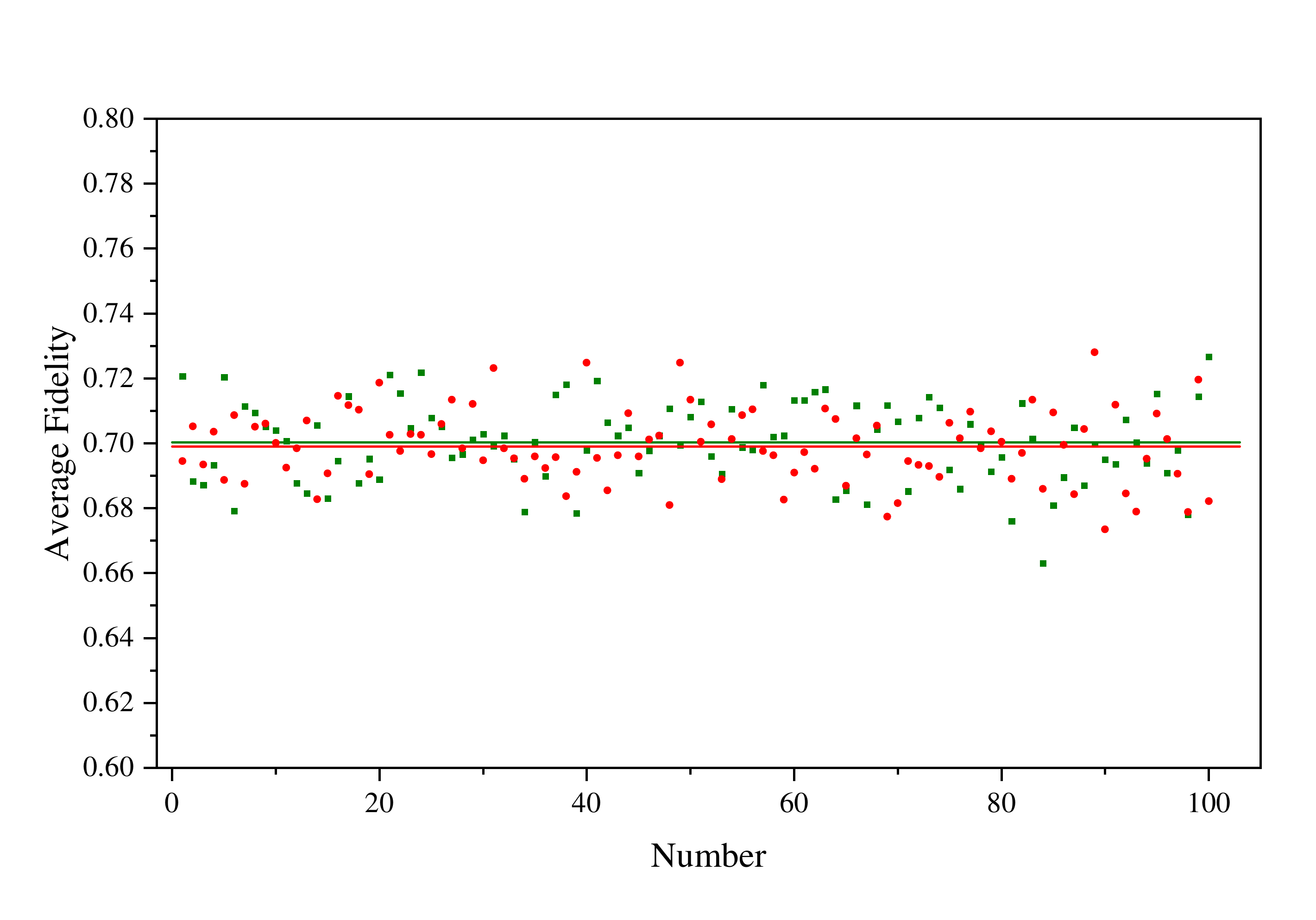}
\end{center}
\caption{Fidelity of 12 MUB state teleported states in MUBs and non-MUBs reconstruction process matrices. The red dots represent the average fidelity of the 12 teleported states obtained by reconstructing the process matrix using the non-MUBs. The green dots represent the average fidelity of the 12 teleported states obtained by reconstructing the process matrix using the MUBs. We repeated it 100 times, with the red line and the green line representing the average of this 100 value using MUBs and non-MUBs, respectively. This means that the two reconstruction methods have little effect on our experimental results.}
\label{fig15}
\vspace{-0.5cm}
\end{figure*}

We convert these probabilities without statistical fluctuation into photon numbers at our experiment counting rate (total 150), and then use Poisson distribution to simulate the fluctuation of data. Then, we use these data to reconstruct the process matrix with two methods. In order to simulate the teleportation process using two reconstruction methods, we evolve the ideal 12 MUB states in the two process matrices to calculate the average fidelity of the teleported state with the ideal state. As shown in Fig.~\ref{fig15}, we repeat this simulation experiment 100 times, there are statistical fluctuations in these 100 average fidelity. The average of these 100 values is very close (for MUB is $0.700\pm0.001$ and for non-MUB is $0.699\pm0.001$). This proves that the use of MUBs or non-MUBs have little effect on our experimental result.

\end{document}